\documentclass[onecolumn]{article}

\usepackage{amsmath,amssymb,amsfonts,bm, slashed,graphicx,longtable}
\usepackage{array}
\usepackage{jheppub}

\newcommand{\amB}[2]{(#1| #2)}
\newcommand{\sml}[1]{\mbox{\scriptsize{$#1$}}}
\newcommand{\Acp}{\mathcal{A}_{\rm CP}}

\newcounter{incre}
\newcommand{\benum}{\begin{enumerate}\setcounter{enumi}{\value{incre}}}
\newcommand{\eenum}{\setcounter{incre}{\value{enumi}}\end{enumerate}}

\usepackage{xcolor}
\definecolor{red}{rgb}{1.0, 0, 0}

\begin{document}

\title{SU(3) Sum Rules for Charm Decay}
\author{Yuval Grossman}
\author{Dean J. Robinson}
\emailAdd{yg73@cornell.edu}
\emailAdd{djr233@cornell.edu}
\affiliation{Laboratory for Elementary-Particle Physics, Cornell University, Ithaca, N.Y.}

\abstract{
We present flavor SU(3) sum rules for $D \to PP$ and $D \to PV$ decay amplitudes, that are valid to second order in symmetry breaking by the strange quark mass spurion. Decay rate sum rules are also computed to this order. Particular attention is given to sum rules arising from the isospin and U-spin subgroups, the former providing sensitive tests for alternative sources of SU(3) breaking. We apply the latter together with the postulated $\Delta U = 0$ rule for the large penguin picture to predict the ratio and difference of the direct CP asymmetries for $D \to KK^*$ and $D \to \pi \rho$.
}

\maketitle
\setcounter{page}{2}

\section{Introduction}
The origin of the unexpectedly large direct CP asymmetry $\Delta\Acp \equiv \Acp[D^0 \to K^-K^+] - \Acp[D^0\to \pi^-\pi^+]$ \cite{Staric:2007dt,Staric:2008rx,Aubert:2007en,Aubert:2007if,Aaij:2011in} is yet to be explained. Any explanation of this result relying upon new physics \cite{Wang:2011uu,Chen:2012am,Giudice:2012qq,Bediaga:2012py,Isidori:2011qw,Chang:2012gna,Grossman:2006jg,Altmannshofer:2012ur,Hochberg:2011ru,Feldmann:2012js,Isidori:2012yx,Grossman:2012eb}  must first properly determine the contribution from the Standard Model (SM). To this end, several studies \cite{Golden:1989ec,Buccella:1994nf,Brod:2012ud,Li:2012cfa,Franco:2012ck,Pirtskhalava:2011va,Bhattacharya:2012kq,Cheng:2012xb} have produced consistent pictures in which the large CP asymmetry can solely originate in the SM, via enhancement of the penguin amplitudes. These studies exploit the approximate flavor SU(3) or U-spin symmetries and show that they admit patterns of penguin enhancement consistent with experimental results. In some approaches \cite{Pirtskhalava:2011va,Bhattacharya:2012kq,Cheng:2012xb}, it can be shown that enhanced penguins can be consistently globally fitted to the data, to first order in the flavor SU(3) breaking.  Moreover, it can be shown that large non-perturbative `penguin contraction' final state interactions  \cite{Brod:2012ud,Li:2012cfa,Golden:1989ec,Buccella:1994nf,Franco:2012ck} can sufficiently enhance the penguins, such that the data can be explained. One approach in particular \cite{Brod:2012ud} has demonstrated that penguin contraction contributions to the $\Delta U =0$ penguin reduced matrix elements -- the so-called $\Delta U = 0$ rule for large broken penguins -- yields a consistent picture for the U-spin subgroup irreps.

Explanations of the direct CP asymmetry excess by particular patterns of flavor SU(3) breaking (hereafter just SU(3), unless otherwise indicated) are complicated by the simultaneous empirical observation of both large SU(3) breakings \emph{and} SU(3) sum rules. Generically, one expects the scale of SU(3) (or U-spin) breaking to be comparable to the splitting of the kaon and pion decay constants, i.e., 
\begin{equation}
\varepsilon \equiv f_K/f_\pi -1 \sim 0.2~,
\end{equation} 
and therefore all SU(3) relations are expected to be violated at this order. However, measuring the reduced square amplitude, defined to be
\begin{equation}
	\label{eqn:DABR}
	|\amB{D}{f}|^2 \equiv \Gamma[D \to f]m_D^2/p_f~,
\end{equation}
in which $p_f$ is the center-of-mass momentum of the final state, one finds empirically the Cabibbo-weighted amplitude relation
\begin{equation}
	\label{eqn:KPR}
	\bigg|\frac{\amB{D^0}{K^+K^-}/V_{cs}^*V_{us}^{\phantom{*}}}{\amB{D^0}{\pi^+\pi^-}/V_{cd}^*V_{ud}^{\phantom{*}}}\bigg| - 1  = 0.82 \pm 0.02~,
\end{equation}
together with the U-spin sum rule
\begin{equation} 
	\label{eqn:USSR}
	\frac{|\amB{D^0}{K^+K^-}/V_{cs}^*V_{us}^{\phantom{*}}| + |\amB{D^0}{\pi^+\pi^-}/V_{cd}^*V_{ud}^{\phantom{*}}|}{|\amB{D^0}{K^+\pi^-}/V_{cd}^*V_{us}^{\phantom{*}}| + |\amB{D^0}{\pi^+K^-}/V_{cs}^*V_{ud}^{\phantom{*}}|} - 1 = 0.040 \pm 0.016~.
\end{equation}
That is, the former is comparable to $\mathcal{O}(1)$ and the latter to $\mathcal{O}(\varepsilon^2)$, rather than the expected $\mathcal{O}(\varepsilon)$.  

Sum rules such as eq. (\ref{eqn:USSR}) are actually a generic consequence of flavor SU(3) breaking. They may exist to arbitrary orders in the SU(3) breaking, although there may be no such sum rules once the order of breaking is sufficiently high, depending on the pattern of symmetry breaking. Commonly, one assumes large SU(3) breaking by the spurion associated with the strange quark mass (see e.g. \cite{Savage1991:su,Hinchliffe:1995hz, Brod:2012ud,Pirtskhalava:2011va,Bhattacharya:2012kq}). Hereafter we call this spurion the s-mass spurion. For example, in the $\Delta U = 0$ rule approach \cite{Brod:2012ud}, certain enhanced U-spin breaking penguins significantly contribute to relations such as eq. (\ref{eqn:KPR}) or $\Delta\Acp$, but the particular U-spin sum rule (\ref{eqn:USSR}) is preserved under the s-mass spurion pattern of breaking to $\mathcal{O}(\varepsilon^2)$, yielding a consistent picture of the experimental results. 

One can naturally extend the s-mass spurion breaking pattern to the full SU(3). An immediate programme is to find the consequent sum rules, which compared to  (\ref{eqn:USSR}) involve the many other D meson decay modes that furnish the SU(3) irreps. Verifying such sum rules is a generic test of any picture of charm decays that invokes this pattern of SU(3) breaking. In this paper we compute the SU(3) sum rules that are valid to $\mathcal{O}(\varepsilon^2)$ in the SU(3) breaking by the s-mass spurion. We further compute the square amplitude sum rules to this order, which have the added advantage of not depending on strong phases. We call these rate sum rules, due to their dependence only on decay rates.  Particular attention is given to sum rules which arise from isospin or U-spin. The former are expected to have parametrically smaller breakings, providing sensitive tests of alternate source of SU(3) breaking. The latter produce square amplitude sum rules to $\mathcal{O}(\epsilon^2)$, and are therefore easier to verify. Where feasible, we shall also discuss current experimental verification of these broken SU(3) sum rules, or predictions arising from them.

This paper is structured as follows. We first briefly recapitulate the construction of the $D$ meson decay amplitudes in terms of reduced matrix elements using the Wigner-Eckhart theorem, and the decomposition of the effective Hamiltonian into SU(3) irreps.  We then proceed to compute the $D \to PP$ and $D \to PV$ amplitudes -- $P$ ($V$) denotes pseudoscalar (vector) -- in terms of their reduced matrix elements to $\mathcal{O}(\varepsilon^2)$, explicit results being provided in appendices.  In doing so, we emphasize that unlike Refs. \cite{Savage1991:su,Hinchliffe:1995hz,Pirtskhalava:2011va,Bhattacharya:2012kq} we do not assume SU(3) breaking arises only from the lowest SU(3) irreps, nor do we neglect doubly Cabbibo-suppressed (DCS) amplitudes. From these results, we extract both amplitude and rate sum rules, valid to $\mathcal{O}(\varepsilon^2)$. We briefly discuss current experimental measurements of the novel sum rules, and proceed to predict ratio and difference of the direct CP asymmetries for $D \to K K^*$ and $D \to \pi\rho$ under the $\Delta U = 0$ rule \cite{Brod:2012ud}. We also show in an appendix how to derive the zeroth order sum rules without computing the reduced matrix elements explicitly.

\section{Framework}

\subsection{Amplitudes and Notation}
We write the in-state D-meson SU(3) triplet and out-state pseudoscalar and vector SU(3) octets and singlets in the usual tensor coefficient notation
\begin{align}
	\label{eqn:DPVM}
	[D_3]^i & =   \begin{pmatrix}  D^0 \\  D^+ \\  D^+_s \end{pmatrix}~,\qquad [P_1] = \eta_1~, \qquad [V_1] = \phi_1~,\notag\\[3pt]
	[P_8]^i_j &  = \begin{pmatrix} \frac{1}{\sqrt{2}}\pi_0 + \frac{1}{\sqrt{6}}\eta_8 & \pi^+ & K^+ \\ \pi^- & - \frac{1}{\sqrt{2}}\pi_0	+ \frac{1}{\sqrt{6}}\eta_8  & K^0 \\ K^- & \bar{K}^0 & -\sqrt{\frac{2}{3}}\eta_8\end{pmatrix}~,\notag\\[3pt] 
	[V_8]^i_j &   = \begin{pmatrix} \frac{1}{\sqrt{2}}\rho_0 + \frac{1}{\sqrt{6}}\omega_8 & \rho^+ & K^{*+} \\ \rho^- & - \frac{1}{\sqrt{2}}\rho_0 + \frac{1}{\sqrt{6}}\omega_8  & K^{*0} \\ K^{*-} & \bar{K}^{*0} & -\sqrt{\frac{2}{3}}\omega_8\end{pmatrix}~.\notag\\
\end{align}
Hereafter Latin indices are SU(3) tensor indices, while Greek indices label a particular state, so that for $M \in \{D_3, P_8, P_1, V_8, V_1\}$, then $(M_\alpha)^i_j = \partial M^i_j /\partial M_\alpha$ is the tensor corresponding to state $M_\alpha$.

In general, for a Hamiltonian $H$ -- presumed to be an SU(3) tensor operator -- we are interested in constructing decay amplitudes of the form
\begin{equation}
	\label{eqn:AD}
	A_{\mu \to \alpha\beta} \equiv \big\langle M_\alpha N_\beta \big|H\big| [D_3]_\mu \big\rangle~.
\end{equation} 	
The Wigner-Eckhart theorem ensures that	
\begin{equation}
	\label{eqn:WET}
	A_{\mu \to \alpha\beta}  =  \sum_w X_w (C_w)_{\alpha\beta\mu}~,\quad  (C_w)_{\alpha\beta\mu}   = \frac{\partial^3}{\partial M _\alpha\partial N _\beta\partial [D_3]_\mu}\bigg[M^i_j N^k_lH^{p_1\cdots p_n}_{q_1\cdots q_m} [D_3]^r\bigg]_w~.
\end{equation}
Here the square brackets indexed by $w$ denote a linearly independent contraction of the SU(3) indices, $X_w$ is the reduced matrix element for each such contraction, $M$, $N \in \{P_{1,8}, V_{1,8}\}$, and $H^{p_1\cdots p_n}_{q_1\cdots q_m}$ are the tensor components of the effective Hamiltonian.  Each contraction $C_w$ is a Wigner-Eckhart invariant, and note that eq.~(\ref{eqn:WET}) implies the Hamiltonian can be written as
\begin{equation}
	\label{eqn:HXC}
	H = \sum_wX_wC_w~.
\end{equation}
The amplitudes $A_{\mu \to \alpha\beta}$ are therefore fully specified by partial derivatives of the Wigner-Eckhart invariants and the reduced matrix elements. Note that in the case that $M_\alpha = N_\beta$, the partial derivatives in (\ref{eqn:WET}) naturally encode an extra factor of $2$, which is the expected combinatoric factor. However, in comparison to the reduced amplitude $\amB{D}{M_\alpha M_\alpha}$ defined in eq. (\ref{eqn:DABR}), we have for mass eigenstates $M_\alpha$
\begin{equation}	
	\label{eqn:AMB}
	\amB{D}{M_\alpha M_\alpha} = \frac{1}{\sqrt{2}}A_{D \to M_\alpha M_\alpha}
\end{equation}
due to the symmetry factor of $1/2$ appearing in the decay rate. 

\subsection{Effective Electroweak Hamiltonian}
In the SM, $\Delta C = -1$ decays arise at leading order from an effective electroweak Hamiltonian with respectively tree and penguin terms of form \cite{Buras:1998raa}
\begin{equation}
	\label{eqn:WEH}
	\frac{G_F}{\sqrt{2}} V^{\phantom{*}}_{u q_1} V^*_{c q_2} (\bar{u}q_1)_L(\bar{q}_2 c)_L~,  \quad -\frac{G_F}{\sqrt{2}}  V^{\phantom{*}}_{u b} V^*_{c b} (\bar{q} q)_{L,R}(\bar{u} c)_L~,
\end{equation}
in which $q_{1,2}$ ($\bar{q}_{1,2}$) are (anti)-quark operators $u,d,s$ as appropriate and $V$ is the Cabbibo-Kobayashi-Maskawa (CKM) matrix. The brackets denote Lorentz and color structure, such that 
$(\bar{q}_1q_2)_{L,R} \equiv (\bar{q}_{1a})_{L,R} \gamma_\mu (q_2^{b})_{L,R}$, with color indices $a$ and $b$ contracted either together or with the adjacent bracket. That is, the operator 
\begin{equation}
	(\bar{q}_1q_2)(\bar{q}_3q_4) = C_1 (\bar{q}_{1a} q_2^b)(\bar{q}_{3b}q_4^a) + C_2 (\bar{q}_{1a} q_2^a)(\bar{q}_{3b}q_4^b)~,
\end{equation}
where $C_i$ are Wilson coefficients, and the former color contraction arises from the color SU(3) completeness relation applied to QCD final or initial state interactions. Hereafter we drop the chiral labels $L$ and $R$, as they are implied by context.

In the SU(3) picture, the operators (\ref{eqn:WEH}) embed into the SU(3) four-quark Hamiltonian, which is the tensor operator
\begin{equation}
	H =H^{k}_{ij}(\bar{q}^{i}q_k)(\bar{q}^jc)~.
\end{equation}
This tensor decomposes as $\bm{\bar{3}}\otimes\bm{3}\otimes\bm{\bar{3}} = \bm{\bar{3}_p} \oplus \bm{\bar{3}_t} \oplus \bm{6} \oplus \bm{15}$. Adopting the tensor coefficient notation $H_{ij}^k \equiv (\bar{q}_i q^k) (\bar{q}_j c)$, one finds explicitly the decomposition
\begin{equation}
	\label{eqn:DHR}
	H_{ij}^k = \delta^k_j\bigg(\frac{3}{8}[\bm{\bar{3}_t}]_i - \frac{1}{8}[\bm{\bar{3}_p}]_i\bigg) + \delta^k_i\bigg(\frac{3}{8}[\bm{\bar{3}_p}]_j - \frac{1}{8}[\bm{\bar{3}_t}]_j\bigg) + \varepsilon_{ijl} [\bm{6}]^{lk} + [\bm{\bar{15}}]_{ij}^{k}~,
\end{equation}
in which the QED-preserving independent components of the $H$ irreps are
\begin{align}
	[\bm{\bar{3}_p}]_1 & = (\bar{u}u)(\bar{u}c) + (\bar{d}d)(\bar{u}c) + (\bar{s}s)(\bar{u}c)\notag\\
	[\bm{\bar{3}_t}]_1 & = (\bar{u}u)(\bar{u}c) + (\bar{u}d)(\bar{d}c) + (\bar{u}s)(\bar{s}c)\notag\\
	[\bm{6}]^{22} & = \frac{1}{2}[(\bar{s}d)(\bar{u}c) - (\bar{u}d)(\bar{s}c)]\notag\\
	[\bm{6}]^{23} & = \frac{1}{4}[(\bar{u}d)(\bar{d}c)-(\bar{d}d)(\bar{u}c) + (\bar{s}s)(\bar{u}c) - (\bar{u}s)(\bar{s}c)]\notag\\
	[\bm{6}]^{33} & = \frac{1}{2}[(\bar{u}s)(\bar{d}c) - (\bar{d}s)(\bar{u}c)]\notag\\
	[\bm{\bar{15}}]^3_{12} & = \frac{1}{2}[(\bar{u}s)(\bar{d}c) + (\bar{d}s)(\bar{u}c)]\notag\\
 	[\bm{\bar{15}}]^2_{13} & =  \frac{1}{2}[(\bar{s}d)(\bar{u}c) + (\bar{u}d)(\bar{s}c)]\notag\\
	[\bm{\bar{15}}]^2_{12} & = \frac{3}{8}[(\bar{u}d)(\bar{d}c) + (\bar{d}d)(\bar{u}c)] - \frac{1}{4}(\bar{u}u)(\bar{u}c) - \frac{1}{8}[(\bar{u}s)(\bar{s}c) + (\bar{s}s)(\bar{u}c)]\notag\\
	[\bm{\bar{15}}]^3_{13} & = \frac{3}{8}[(\bar{u}s)(\bar{s}c) + (\bar{s}s)(\bar{u}c)] - \frac{1}{4}(\bar{u}u)(\bar{u}c) - \frac{1}{8}[(\bar{u}d)(\bar{d}c) + (\bar{d}d)(\bar{u}c)]~.\label{eqn:SU3DH}
\end{align}	
All other components are set to zero due to charge conservation. Eqs. (\ref{eqn:WEH}) imply that the tensor components of the electroweak Hamiltonian may be obtained at leading order from the map $(\bar{u}q_1)(\bar{q}_2c) \mapsto V^{\phantom{*}}_{u q_1} V^*_{cq_2}$, $(\bar{q}q)(\bar{u}c) \mapsto -V^{\phantom{*}}_{u b} V^*_{cb}$ and other terms zero. Unitarity of the CKM matrix and its Wolfenstein parametrization yields finally the independent $H$ components, to leading order in $\lambda$
\begin{gather}
	[\bm{\bar{3}_p}]_1 \simeq -2\lambda^5A^2(\rho - i\eta)~, \qquad  [\bm{\bar{3}_t}]_1 \simeq  -\lambda^5A^2(\rho - i\eta)~, \notag\\
	[\bm{6}]^{22}  \simeq -\frac{1}{2}~, \qquad [\bm{6}]^{23}  \simeq -\frac{\lambda}{2}~, \qquad [\bm{6}]^{33}  \simeq -\frac{\lambda^2}{2}~,\notag\\
	[\bm{\bar{15}}]^3_{12}  \simeq -\frac{\lambda^2}{2}~, \qquad [\bm{\bar{15}}]^2_{13}  \simeq  \frac{1}{2}~, \qquad [\bm{\bar{15}}]^2_{12}  \simeq -\frac{\lambda}{2}~, \qquad [\bm{\bar{15}}]^3_{13}  \simeq \frac{\lambda}{2}~. \label{eqn:CKMH}
\end{gather}
It is apparent from the CKM structure that $\bm{\bar{3}_{p,t}}$ will produce penguin-like contributions to an amplitude, with a CP violating phase -- i.e. $\propto \lambda^5A^2(\rho - i\eta)$ --  while the $\bm{6}$ and $\bm{\bar{15}}$ produce tree-like CF, SCS and DCS terms. 

Finally, note that in this parameterization the two $\bm{\bar{3}}$ irreps of eq. (\ref{eqn:DHR}) are linear combinations of $\bm{\bar{3}_{p,t}}$ and hence always proportional to one another at leading order in $\lambda$; they are not linearly independent. This means we need only consider a single $\bm{\bar{3}}$ when computing amplitudes from the invariants and reduced matrix elements. Henceforth, without loss of generality we consider just $\bm{\bar{3}_p}$ for this purpose, multiplying it by a factor of 3/8 to match the first $\bm{\bar{3}}$ irrep of eq. (\ref{eqn:DHR}), and we hereafter call the resulting irrep simply $\bm{\bar{3}}$.

\subsection{SU(3) Breaking}
Under the assignment of eqs. (\ref{eqn:CKMH}), the electroweak Hamiltonian, $H$, itself may be thought of as an SU(3) violating spurion.  We assume further SU(3) breaking is produced by the $s$-mass spurion, which in traceless tensor form is
\begin{equation}
	m_s = \varepsilon\begin{pmatrix} ~1 & ~0 & ~0 \\ ~0 & ~1 & ~0 \\ ~0 & ~0 & -2\end{pmatrix}~, \quad \varepsilon \sim 0.2~.
\end{equation}
The Hamiltonian becomes $H + Hm_s$ at first order in the spurion, i.e. at order $\mathcal{O}(\varepsilon)$. By eq. (\ref{eqn:WET}) the corresponding amplitudes are
\begin{equation}
	A_{\mu \to \alpha\beta} = \langle M_\alpha M_\beta |H|D_\mu\rangle +  \langle M_\alpha M_\beta |H m_s |D_\mu\rangle \equiv \sum_w X_w(C_w)_{\alpha\beta\mu} + \varepsilon \sum_w X_{w,s}(C_{w,s})_{\alpha\beta\mu}~,
\end{equation}
the subscript `$s$' denoting the first order s-mass spurion contributions.  Since we expect $X_{w,s} \sim \mathcal{O}(1)$, then corrections arising from the $n$th order $H m_s^n$ spurion term are expected to be $\mathcal{O}(\varepsilon^n)$. 

A second, parametrically smaller, source of SU(3) breaking arises from the $u$-$d$ mass splitting. That is, isospin breaking due to the spurion
\begin{equation}
	m_I = \delta\begin{pmatrix} ~1 & ~0 \\ ~0 & -1 \\ \end{pmatrix}~,\quad \delta = (m_u - m_d)/\Lambda_{\rm qcd} \sim 1\%~,
\end{equation}
which we have written in the adjoint representation of the isospin subgroup, rather than as a SU(3) tensor. This spurion similarly introduces $Hm_I^n$ corrections at $\mathcal{O}(\delta^n)$, the first order correction being $\delta\sum X_{w,I} C_{w,I}$, whose invariants can be computed similarly to those of $m_s$.

In this language, the key idea of the large broken penguin picture is that certain $X_w$ and $X_{w,s}$ are enhanced. For example, under the $\Delta U = 0$ rule of Ref.~\cite{Brod:2012ud}, the reduced matrix elements associated with exclusively $\Delta U = 0$ operators are enhanced. One might propose an extension of this rule to the SU(3) picture, which would enhance the reduced matrix elements associated with contractions involving the $\Delta U = 0$ components $[\bm{\bar{3}}]_1$, $[\bm{6}]^{23}$, $[\bm{\bar{15}}]^{2}_{12}$ and $[\bm{\bar{15}}]^{3}_{13}$.  We will discuss the implications of this idea briefly below, but a full global fit of this proposed picture to the data is beyond the scope of the present paper.

\subsection{Formal Sum Rules}
A key feature of SU(3) breaking, and the focus of this paper, is the set of associated sum rules, which can be computed to arbitrary order in $\varepsilon$. To be precise, a sum rule is a symbol $\mathcal{S}$, such that
\begin{equation}
	\label{eqn:DSR}
	\mathcal{S}^{\alpha\beta\mu}\mathcal{A}_{\mu \to \alpha\beta} = 0 
\end{equation}
which is equivalent to
\begin{equation}
	\label{eqn:DSRC}
	 \mathcal{S}^{\alpha\beta\mu}(C_w)_{\alpha\beta\mu} = 0~, ~~\forall w~,
\end{equation}
noting that $w$ labels the invariants. In general, a sum rule may be found to $\mathcal{O}(\varepsilon^n)$ by computing the appropriate invariants to that order, and solving the linear equations (\ref{eqn:DSRC}), that is, finding the kernel of $(C_w)_{\alpha\beta\mu}$. The number of sum rules is a non-increasing function of $n$, and the number of sum rules may be zero once $n$ is sufficiently high, depending on the pattern of symmetry breaking.

Alternatively, as we show in Appendix~\ref{app:ASRG}, the symmetries of the Hamiltonian $H$ may be sometimes used to compute sum rules directly, without needing to first compute the invariants. The key idea is that if there exists an operator $T$ under which $H$ is invariant, that is $TH = 0$, then it follows that 
\begin{equation}
	\label{eqn:FS}
	T^{\rho\sigma\gamma}_{\alpha\beta\mu}(C_{w})_{\rho\sigma\gamma} = 0~,
\end{equation}
where the indices here are the indices of the corresponding final and initial state irreps, rather than tensor indices. For example, if $T$ is an operator that changes electric charge by $\Delta Q$, then choosing $\alpha$, $\beta$ and $\mu$ corresponding to an amplitude which violates QED by  $-\Delta Q$ produces a sum rule of QED preserving amplitudes. That is $\mathcal{S} = T_{\alpha\beta\mu}$. In Appendix~\ref{app:ASRG} we compute the zeroth order in $m_s$ sum rules for $D \to PP$ by this alternate method.

On a presentational note, we emphasize that a linear combination of sum rules is also a sum rule, so that there is an arbitrarily large way to write any set of $m$ linearly independent sum rules. In particular, in finding the kernel of $C_w$, one may often find short (long) sum rules involving a small (large) number of amplitudes. In writing the sum rules, we have adopted the preference that the longest sum rules have as minimal length as possible, while well-known sum rules, in particular the U-spin and the isospin sum rules, are also explicitly presented. There exists no algorithm that achieves this preference. Instead we have employed an approximate computational method in which the $m$-dimensional kernel is computed repeatedly under random permutations of the amplitude basis, forming a large list of dependent sum rules. After extracting well-known sum rules from this list, we then extract the shortest remaining linearly independent set of sum rules that will span -- together with the well-known ones -- the $m$-dimensional kernel.

\subsection{Rate Sum Rules}
Amplitude sum rules can only be verified if the strong phases of each amplitude are known. Sum rules involving square amplitudes -- rate sum rules -- are therefore particularly useful, since these correspond to sum rules of branching ratios or decay rates, provided the corresponding phase spaces are not zero. We may similarly compute the rate sum rules by observing that
\begin{equation}
	\label{eqn:SASR}
	\big| \mathcal{A}_{\mu \to \alpha\beta} \big|^2 = \sum_w \sum_{w'} X_w X^*_{w'} (C_w)_{\alpha\beta\mu}(C_{w'})^*_{\alpha\beta\mu} \equiv \sum_u X_u (C_u)_{\alpha\beta\mu}
\end{equation}
where $u = \{w,w'\}$, $X_u = X_w X_{w'}^*$ and $C_u = C_w C_{w'}$;  $C_w$ are real. Eq. (\ref{eqn:SASR}) implies that one need only solve the linear equations $\mathcal{S}^{\alpha\beta\mu}(C_u)_{\alpha\beta\mu} = 0$,  $\forall u$ in order to obtain the sum rules at the desired order.  

\subsection{Isospin and U-spin Sum Rules}
\label{sec:SSISR}
So far we have embedded the electroweak Hamiltonian into flavor SU(3) irreps, such that it is an SU(3) spurion. The effective Hamiltonian itself then generates invariants and SU(3) sum rules; a subspace of these sum rules holds once $n$th order s-mass spurion terms are added. We may, however, alternatively embed the effective Hamiltonian into SU(3) subgroups, in particular U-spin and isospin, and construct invariants involving only those initial and final states which furnish irreps of these subgroups. The Hamiltonian is then a spurion of these subgroups, and in this manner we may obtain sum rules, which we call the U-spin or isospin sum rules respectively. Of course, these isospin and U-spin sum rules necessarily also arise in the full SU(3) picture itself, but generally in linear combinations with other pure SU(3) sum rules.

Returning to the full SU(3) picture, while the electroweak Hamiltonian itself necessarily breaks isospin, observe the s-mass spurion $m_s$ does not; it breaks SU(3) down to isospin $\times$ strangeness (or equivalently QED). One therefore expects isospin sum rules obtained from the electroweak Hamiltonian itself to be preserved to all orders in $m_s$, even though isospin itself is already broken by $H$. One can verify this expectation explicitly with the language of eq. (\ref{eqn:FS}): In the isospin picture, if $T$ is an isospin tensor operator that generates isospin sum rules, i.e. $TH = 0$, then since $m_s$ is an isospin singlet, $Tm_s = 0$. This result naturally embeds into the full SU(3) picture, thereby showing that isospin sum rules are unbroken to all orders in $m_s$.

This result is particularly useful: If there is no other significant source of SU(3) breaking other than $m_s$ and $m_I$, then isospin sum rules valid to second order in the \emph{isospin} spurion $m_I$ -- i.e. valid to $\mathcal{O}(\delta^2)$ -- are expected to hold to the $10^{-4}$ level. Verifying such a sum rule is therefore an extremely sensitive test of the presence of other sources of SU(3) breaking, which includes new physics. We shall examine such sum rules for the $D\to PP$ and $D \to PV$ case. It should be noted that sensitive sum rule tests based on isospin have previously been proposed for charmless B decays \cite{Gronau:2006eb}, although these depended on dynamical suppression of first order isospin breaking. Here the effect is due to the pattern of symmetry breaking itself.
 
\subsection{Mixing}
\label{sec:FM}
The states furnishing the SU(3) octet and singlets do not always correspond to meson mass eigenstates. In particular, one must account for $K-K$, $\omega-\phi$ and $\eta-\eta'$ mixing. In the CP limit, which we assume for kaon mixing, the mixing of $K^0-\bar{K}^0$ is maximal, so we define the usual mass eigenstates
\begin{equation}
	\label{eqn:KM}
	\begin{pmatrix} K_S \\ K_L \end{pmatrix} = \frac{1}{\sqrt{2}}\begin{pmatrix}  \phantom{-}1 &  \phantom{-}1 \\ -1 & \phantom{-}1 \end{pmatrix}\begin{pmatrix} K^0 \\ \bar{K}^0 \end{pmatrix}~.
\end{equation}
Similarly, the $\omega$-$\phi$ mixing is idealized such that the $\phi$ mass eigenstate is pure $\bar{s}s$. I.e. the mass eigenstates
\begin{equation}
	\begin{pmatrix} \omega \\ \phi \end{pmatrix} = \frac{1}{\sqrt{3}}\begin{pmatrix} 1 & \sqrt{2} \\ -\sqrt{2} & 1 \end{pmatrix}\begin{pmatrix} \omega_8 \\ \phi_1 \end{pmatrix}~.
\end{equation}
Finally, in the case of $\eta-\eta'$ mixing, the mixing angle is neither ideal nor maximal \cite{Thomas:2007uy,Ambrosino:2009sc,Cao:2012nj}, so one defines mass eigenstates
\begin{equation}
	\begin{pmatrix} \eta \\ \eta' \end{pmatrix} = \begin{pmatrix} \cos\theta & -\sin\theta \\ \sin\theta & \cos\theta \end{pmatrix}\begin{pmatrix} \eta_8 \\ \eta_1 \end{pmatrix}~.
\end{equation}
Hereafter while the invariants are computed in terms of the flavor SU(3) states, we shall write the sum rules in terms of either the mass or flavor basis, depending on convenience.

\section{\texorpdfstring{$D \to PP$}{DPP} Sum Rules}
In this section we present the $D \to PP$ amplitude and rate sum rules that are valid to $\mathcal{O}(\varepsilon^2)$, that is, they are na\"\i vely broken at $\mathcal{O}(\varepsilon^2)$. The corresponding SU(3) invariants are presented in detail in Appendix \ref{app:PP}. We emphasize that we compute the sum rules only from tree-like operators. That is, we neglect the sub-leading invariants of order $\mathcal{O}(\lambda^5)$ due to the $\bm{\bar{3}}$ irreps, and consider only the invariants produced by the $\bm{6}$ and $\bm{\bar{15}}$.

We note in passing that the SU(3) invariants constructed from the $\bm{6}$ and $\bm{\bar{15}}$ generally involve both $\Delta U = 0$ and $\Delta U = \pm 1$ operators, but we see from the tables that only $\Delta U = 0$ amplitudes receive corrections from
\begin{equation}
	C_{[_{88}^{\rm pp}]_{\bm{6}s}^{2}}~,\qquad C_{[_{88}^{\rm pp}]_{\bm{\bar{15}}s}^1}~,\qquad \mbox{and}\qquad C_{[_{88}^{\rm pp}]_{\bm{\bar{15}}s}^7}~.
\end{equation}
That is, these invariants must involve exclusively $\Delta U = 0$ operators. A $\Delta U = 0$ rule could then be implemented by enhancing only those reduced matrix elements associated with these three invariants. However, we do not consider this possibility further in the present work.

\subsection{\texorpdfstring{$P_1P_8$}{OPP} Amplitude Sum Rules}
Sum rules valid at first order in the spurion can be extracted from the invariants presented in Appendix~\ref{app:PP}.  One finds that there are twelve sum rules for $P_8P_8$ and two for $P_1P_8$. The two $P_1P_8$ sum rules are, in the SU(3) flavor basis

\setcounter{incre}{0}

\benum
 \item 
\begin{align*}
 -\frac{A_{D^+\to \eta_1  K^+}}{\lambda^2 }& +\frac{A_{D^+\to \eta_1  \pi ^+}}{\lambda}-\frac{A_{D^+_s\to \eta_1  K^+}}{\lambda}+A_{D^+_s\to \eta_1  \pi ^+} =0
\end{align*}
 \item
 \begin{align*}
   - \frac{\sqrt{3}A_{D^0\to \eta_1  \eta_8 }}{\lambda}&+\frac{\sqrt{2} A_{D^0\to \eta_1  K^0}}{\lambda^2 }+\frac{A_{D^0\to \eta_1  \pi _0}}{\lambda}-\sqrt{2}   A_{D^0\to \eta_1  \bar{K}^0} =0~.
\end{align*}
\eenum

\subsection{Isospin Sum Rules}
For the $P_8P_8$ sum rules, we first identify the pure isospin sum rules. 
Explicitly, note that we have the isospin $\times$ strangeness irreps
\begin{gather}
	\Pi^i_j = \begin{pmatrix} \frac{\pi^0}{\sqrt{2}} & \pi^+ \\ \pi^- & -\frac{\pi^0}{\sqrt{2}} \end{pmatrix} \sim 3_0~, \quad K^i = \begin{pmatrix} K^+ \\ K^0 \end{pmatrix} \sim 2_{-1}~,\quad \eta_{8,1} \sim 1_0~,\notag\\
	 \bar{K}_i = \begin{pmatrix} K^- & \bar{K}^0 \end{pmatrix} \sim 2_{+1}~, \quad D^i = \begin{pmatrix} D^0 \\ D^+ \end{pmatrix} \sim 2_{0}~,\quad D_s^+ \sim 1 _{1}~,\label{eqn:DPII}
\end{gather}
and we expect one sum rule for each of $\Pi \Pi$, $\Pi K$ and $\Pi \bar{K}$ final states. These three sum rules are, 
\benum
\item \label{sum:PPPP}
\begin{align*}
-A_{D^0\to 2\pi_0}& +A_{D^0\to \pi ^- \pi ^+}+\sqrt{2} A_{D^+\to \pi _0 \pi ^+} =0
\end{align*}
\item 
\begin{align*}
 A_{D^0\to \pi _0 \bar{K}^0}& +\frac{A_{D^0\to K^- \pi ^+}}{\sqrt{2}}-\frac{A_{D^+\to \bar{K}^0 \pi ^+}}{\sqrt{2}} =0
\end{align*}
\item \label{sum:PPPK}
\begin{align*}
 \sqrt{2} A_{D^0\to K^0 \pi _0}& +A_{D^0\to \pi ^- K^+}+\sqrt{2} A_{D^+\to \pi _0 K^+}-A_{D^+\to K^0 \pi ^+} =0~.
\end{align*}
\eenum
Note that isospin sum rules must involve amplitudes of the same strangeness violation -- i.e. same $\Delta S$ -- and therefore of the same Cabibbo order, due to the structure of the effective Hamiltonian. As a result $D^0 \to K^0\bar{K}^0$, which occurs only through penguin operators, cannot form a sum rule with the other $K \bar{K}$ processes $D^+ \to K^+\bar{K}^0$ and $D^0 \to K^+ K^-$, which occur at tree level. There is therefore no sum rule for the $K \bar{K}$ final state.

As mentioned in Sec. \ref{sec:SSISR}, these isospin sum rules hold to all s-mass spurion orders, so sum rules valid to second order in isospin breaking are a sensitive test of alternate SU(3) breaking sources. If one computes the invariants at first order in SU(3) breaking by the isospin spurion, $m_I$, one finds that all three isospin sum rules are broken at $\mathcal{O}(\delta)$, as expected. However, the difference sum rule $\mbox{\ref{sum:PPPK}} - \lambda \mbox{\ref{sum:PPPP}} = 0$, i.e. it is valid to $\mathcal{O}(\delta^2)$. Precisely measuring the deviation from zero of the corresponding reduced amplitude sum rule
\begin{equation}
\label{eqn:PTB}
\frac{ \amB{D^+}{ K^0 \pi ^+} -\sqrt{2} \amB{D^+}{ \pi _0 K^+}+\lambda \sqrt{2} \amB{D^+}{ \pi _0 \pi ^+}}{\sqrt{2} \amB{D^0}{ K^0 \pi _0} +\amB{D^0}{ \pi ^- K^+} - \lambda \amB{D^0}{ \pi ^- \pi ^+} + \lambda \sqrt{2}\amB{D^0}{ 2\pi_0} } -1
\end{equation}
could therefore be a comparatively sensitive test of new physics, even if $\Delta \mathcal{A}_{CP}$ itself is due to SM physics. While branching fractions of all seven modes in this sum rule have been measured \cite{pdg:2012}, one must also know the strong phases in order to compute (\ref{eqn:PTB}). Strong phases can be measured from the Dalitz plots of 3-body charm decays (see e.g \cite{Rosner:2003yk}), and determining the strong phases here remains an experimental goal for the future.

\subsection{\texorpdfstring{$P_8 P_8$}{OE} Sum Rules}
\label{sec:PP88SR}
Now, returning to the full SU(3), we choose our sum rule basis such that the above three isospin sum rules are presented explicitly, along with the single U-spin sum rule 
\benum
\item \label{sum:PPUS1}
\begin{align*}
 \frac{A_{D^0\to K^- K^+}}{\lambda }& +\frac{A_{D^0\to \pi ^- K^+}}{\lambda ^2}-A_{D^0\to K^- \pi ^+}-\frac{A_{D^0\to \pi ^- \pi ^+}}{\lambda } = 0~.
\end{align*}
\eenum
The remaining eight linearly independent $P_8P_8$ sum rules are pure SU(3) sum rules, which we find to be
\benum
\item
\begin{align*}
 -\frac{\sqrt{3} A_{D^0\to \eta_8  K^0}}{\lambda ^2}& +\frac{A_{D^0\to K^0 \pi _0}}{\lambda ^2}-\sqrt{3} A_{D^0\to \eta_8  \bar{K}^0}+A_{D^0\to \pi _0 \bar{K}^0} =0
\end{align*}
\item
\begin{align*}
 \frac{\sqrt{3} A_{D^+\to \eta_8  K^+}}{\lambda^2 }& -\frac{A_{D^+\to \pi _0 K^+}}{\lambda^2 }+\sqrt{2} \frac{A_{D^+\to \bar{K}^0 K^+}}{\lambda}+\sqrt{3} 
   A_{D^+_s\to \eta_8  \pi ^+}-\sqrt{2} \frac{A_{D^+_s\to K^0 \pi ^+} }{\lambda}=0
\end{align*}
\item
\begin{align*}
A_{D^+\to \bar{K}^0 \pi ^+}& -A_{D^+_s\to \bar{K}^0 K^+} -\frac{ A_{D^+\to \bar{K}^0 K^+}}{\lambda} +\frac{A_{D^+_s\to K^0
   \pi ^+} }{\lambda} +\frac{A_{D^+\to K^0 \pi ^+}}{\lambda^2}\\[10pt]
   & -\frac{A_{D^+_s\to K^0 K^+} }{\lambda^2}=0
\end{align*}
\item
\begin{align*}
 \sqrt{3} \frac{A_{D^0\to \eta_8  \pi _0}}{\lambda}&-\frac{\sqrt{2} A_{D^0\to K^0 \pi _0}}{\lambda^2 }-\frac{A_{D^0\to 2\pi_0}}{\lambda}-\sqrt{\frac{3}{2}}  A_{D^0\to \eta_8 
   \bar{K}^0}+\frac{A_{D^0\to K^0 \bar{K}^0}}{\lambda}\\[10pt]
   &+\frac{3 A_{D^0\to \pi _0 \bar{K}^0}}{\sqrt{2}} =0
\end{align*}
\item
\begin{align*}
 \frac{A_{D^0\to 2\eta_8}}{\lambda}&+\frac{2 A_{D^0\to \eta_8  \pi _0}}{\sqrt{3}\lambda}-\frac{4 \sqrt{2} A_{D^0\to K^0 \pi _0}}{3 \lambda^2 }-\frac{A_{D^0\to 2\pi_0}}{\lambda}+2 \sqrt{\frac{2}{3}}
  A_{D^0\to \eta_8  \bar{K}^0}\\[10pt]
   & +\frac{2}{3} \sqrt{2}  A_{D^0\to \pi _0 \bar{K}^0} =0
\end{align*}
\item
\begin{align*}
\frac{\sqrt{2} A_{D^+\to \pi _0 K^+}}{\lambda ^2}& -\frac{A_{D^+\to \bar{K}^0 K^+}}{\lambda }-\frac{\sqrt{2} A_{D^+\to \pi _0 \pi ^+}}{\lambda }+A_{D^+\to
   \bar{K}^0 \pi ^+}  +\frac{\sqrt{2} A_{D^+_s\to \pi _0 K^+}}{\lambda }\\[10pt]
   & -A_{D^+_s\to \bar{K}^0 K^+} =0
\end{align*}
\item
\begin{align*}
-\sqrt{\frac{2}{3}}  A_{D^+_s\to \bar{K}^0 K^+}& - \frac {A_{D^+\to \eta_8  \pi ^+}}{\lambda}  -\frac{A_{D^+\to \pi _0 \pi ^+} 
   }{\sqrt{3}\lambda}\\[10pt]
   & +\frac{2  A_{D^+_s\to \pi _0 K^+} }{\sqrt{3}\lambda}+\frac{A_{D^+\to \eta_8  K^+}}{\lambda^2}+\frac{A_{D^+\to \pi _0
   K^+}}{\sqrt{3}\lambda^2}+\sqrt{\frac{2}{3}} \frac{A_{D^+\to K^0 \pi ^+}}{\lambda^2} =0
\end{align*}
\item
\begin{align*}
 \frac{2 A_{D^+\to \pi _0 K^+}}{\sqrt{3} \lambda ^2}& -\sqrt{\frac{2}{3}}\frac{ A_{D^+\to K^0 \pi ^+}}{\lambda ^2}-\frac{2 A_{D^+\to \pi _0 \pi
   ^+}}{\sqrt{3}\lambda}+\frac{A_{D^+_s\to \eta_8  K^+}}{\lambda}+\frac{A_{D^+_s\to \pi _0 K^+}}{\sqrt{3}\lambda} \\[10pt]
   & +\sqrt{\frac{2}{3}}  A_{D^+_s\to \bar{K}^0 K^+}- A_{D^+_s\to \eta_8  \pi ^+}=0~.
\end{align*}
\eenum

Again, these sum rules can only be verified provided the corresponding strong phases for each process can be measured. Conversion to the mass basis is achieved by the relations
\begin{gather}
	A_{D \to f K^0}  = \frac{1}{\sqrt{2}}A_{D \to fK_S} - \frac{1}{\sqrt{2}}A_{D \to fK_L}~, \qquad A_{D \to f \bar{K}^0}  = \frac{1}{\sqrt{2}}A_{D \to fK_S} + \frac{1}{\sqrt{2}}A_{D \to fK_L}~,\notag\\
	A_{D \to K^0\bar{K}^0}  = \frac{1}{2}A_{D \to 2K_S} - \frac{1}{2}A_{D \to 2K_L}~,\notag\\
	A_{D \to f \eta_8}  = \cos\theta A_{D \to f\eta} + \sin\theta A_{D \to f\eta'}~,\qquad A_{D \to f \eta_1}  = -\sin\theta A_{D \to f\eta} + \cos\theta A_{D \to f\eta'}~,\notag\\
\end{gather}
applying the extra symmetry factor (\ref{eqn:AMB}) as appropriate. Note that the amplitudes involving either $\eta_1\eta_8$ or $\eta_8\eta_8$ final states necessarily include a $A_{D \to \eta'\eta'}$ term, which cannot be measured due to its zero phase space. The sum rules including such amplitudes, which here are sum rules (ii) and (xi), therefore cannot be measured from decays. 

\subsection{Rate Sum Rules}
\label{sec:PPRSR}
We next present the rate sum rules valid to $\mathcal{O}(\varepsilon^2)$, which have the added advantage of being directly proportional to the corresponding branching ratios. Following from eq. (\ref{eqn:SASR}), the invariants of the square amplitudes up to and including order $\mathcal{O}(\varepsilon)$ are found by taking all possible $\mathcal{O}(1)$ and $\mathcal{O}(\varepsilon)$ pairwise products of the amplitude invariants in Appendix \ref{app:PP}, taking into account the mixings of Sec. \ref{sec:FM}. Applying eq. (\ref{eqn:AMB}) where appropriate, one finds the following square amplitude sum rules.

\begin{enumerate}
\item
\begin{gather*}
 |\amB{D^+}{ K_L K^+}|^2 = |\amB{D^+}{ K_S K^+}|^2 
\end{gather*}
\item
\begin{gather*}
 |\amB{D^+_s}{ K_L \pi^+}|^2 = |\amB{D^+_s}{ K_S \pi ^+}|^2 
\end{gather*}
\item \label{sum:PPU1}
\begin{gather*}
 \frac{|\amB{D^0}{ K^- K^+}|^2}{\lambda ^2}+\frac{|\amB{D^0}{ \pi ^- \pi ^+}|^2}{\lambda ^2}  = \frac{|\amB{D^0}{ \pi ^- K^+}|^2}{\lambda ^4}+ |\amB{D^0}{ K^- \pi ^+}|^2
 \end{gather*}
\item \label{sum:PPSA1}
\begin{gather*}
  |\amB{D^0}{  2\eta}|^2 +  |\amB{D^0}{ \eta  \pi _0}|^2+ |\amB{D^0}{ 2\pi _0}|^2+  |\amB{D^0}{ \eta  \eta '}|^2 + |\amB{D^0}{ \pi _0 \eta '}|^2+ |\amB{D^0}{ 2\eta'}|^2 \\
  = \\
   \Big[|\amB{D^0}{ \eta  K_L}|^2- |\amB{D^0}{ \eta  K_S}|^2\Big]+  \Big[|\amB{D^0}{ K_L \pi _0}|^2 - |\amB{D^0}{ K_S \pi _0}|^2\Big]\\
 +  \Big[|\amB{D^0}{ K_L \eta '}|^2- |\amB{D^0}{ K_S \eta '}|^2\Big]
\end{gather*}
\item \label{sum:PPSA2}
\begin{gather*}
 \frac{|\amB{D^+}{ \eta  \pi ^+}|^2}{\lambda ^2}+\frac{|\amB{D^+}{ \pi _0 \pi ^+}|^2}{\lambda ^2}+\frac{|\amB{D^+}{ \pi ^+ \eta '}|^2}{\lambda ^2}+\frac{|\amB{D^+_s}{ \eta  K^+}|^2}{\lambda^2} \\
 +\frac{|\amB{D^+_s}{ \pi _0 K^+}|^2}{\lambda ^2} +\frac{|\amB{D^+_s}{ K^+ \eta '}|^2}{\lambda ^2}\\
     =  \\
     \frac{|\amB{D^+}{ \eta  K^+}|^2}{\lambda ^4} + \frac{|\amB{D^+}{ \pi _0 K^+}|^2}{\lambda^4} + \frac{|\amB{D^+}{ K^+ \eta '}|^2}{\lambda^4}+ |\amB{D^+_s}{ \eta  \pi^+}|^2 + |\amB{D^+_s}{ \pi ^+ \eta '}|^2  \\
 +  \bigg[\frac{|\amB{D^+_s}{ K_L K^+}|^2}{\lambda^2}  -\frac{|\amB{D^+_s}{ K_S K^+}|^2}{\lambda ^2}\bigg]+\bigg[\frac{|\amB{D^+}{ K_L \pi^+}|^2}{\lambda ^2} -\frac{|\amB{D^+}{ K_S \pi ^+}|^2}{\lambda ^2}\bigg]
\end{gather*}
\end{enumerate}

The first two sum rules are simply consequences of $K-K$ mixing and sum rule \ref{sum:PPU1} is the U-spin rate sum rule. Combining it with the amplitude U-spin sum rule \ref{sum:PPUS1} of Sec. \ref{sec:PP88SR}, one may show that the normed amplitude sum rule (\ref{eqn:USSR}) is also valid to $\mathcal{O}(\varepsilon^2)$, as expected. The argument to show this proceeds as follows: The rate and amplitude sum rules are a set of equations of the form
\begin{equation}
	|a|^2 + |b|^2 = |c|^2 + |d|^2 + \mathcal{O}(\varepsilon^2)~,\qquad a - b = c - d + \mathcal{O}(\varepsilon^2)~.
\end{equation}
These are satisfied by the relations $a = c + P \varepsilon + \mathcal{O}(\varepsilon^2)$, $b = d + P\varepsilon + \mathcal{O}(\varepsilon^2)$, and $c + d = Q\varepsilon + \mathcal{O}(\varepsilon^2)$, for some $\mathcal{O}(1)$ $P$ and $Q$, which can also be explicitly verified by checking the invariants of Appendix \ref{app:PP}.  The desired relation $|a| + |b| = |c| + |d| + \mathcal{O}(\varepsilon^2)$ -- i.e. eq. (\ref{eqn:USSR})  -- follows. We emphasize that this normed amplitude sum rule is a consequence of the fact that there are both amplitude and rate sum rules involving the same modes, and in the present analysis this circumstance is unique to the U-spin sum rules.

Sum rules \ref{sum:PPSA1} and \ref{sum:PPSA2} are novel to the broken SU(3) picture. Note that sum rule \ref{sum:PPSA1} involves a $D^0 \to 2\eta'$ decay, which has zero phase space. Hence this sum rule is unfortunately not measurable. In contrast, sum rule \ref{sum:PPSA2} is measurable, and requires that the branching ratios and phase space of all these modes be experimentally determined.

\section{\texorpdfstring{$D \to PV$}{DPV} Sum Rules}
As for the $D \to PP$ case, the amplitude and rate sum rules valid to $\mathcal{O}(\varepsilon^2)$ can be extracted from the $D \to PV$ invariants of Appendix \ref{app:PV} by computing the kernel in the amplitude basis.  Once again, we emphasize that in computing the sum rules we neglect the sub-leading invariants of order $\mathcal{O}(\lambda^5)$ due to the $\bm{\bar{3}}$ irreps, and consider only the invariants produced by the $\bm{6}$ and $\bm{\bar{15}}$. 

\subsection{\texorpdfstring{$P_1V_8$}{OPV} and \texorpdfstring{$V_1P_8$}{VP} Amplitude Sum Rules}
The two $P_1V_8$ and two $V_1P_8$ sum rules valid to $\mathcal{O}(\varepsilon^2)$ are respectively in the flavor basis
\setcounter{incre}{0}

\benum
\item
\begin{align*}
 \frac{A_{D^+\to \eta_1 \rho^+}}{\lambda }& -\frac{A_{D^+\to \eta_1 K^{*+}}}{\lambda^2}+A_{D_s^+\to \eta_1 \rho^+}-\frac{A_{D_s^+\to \eta_1 K^{*+}}}{\lambda } =0 
\end{align*}
\item
\begin{align*}
 \frac{\sqrt{3}A_{D^0\to \omega_8  \eta_1}}{\lambda } & -\frac{A_{D^0\to \eta_1 \rho_0}}{ \lambda }-\frac{\sqrt{2} A_{D^0\to \eta_1K^{*0}}}{\lambda^2}+\sqrt{2} A_{D^0\to \eta_1 \bar{K}^{*0}} =0
 \end{align*}
\item
\begin{align*}
 \frac{A_{D^+\to \phi_1  K^+}}{\lambda^2}& -\frac{A_{D^+\to \phi_1  \pi^+}}{\lambda }+\frac{A_{D_s^+\to \phi_1  K^+}}{\lambda}-A_{D_s^+\to \phi_1  \pi^+} = 0
\end{align*}
\item
\begin{align*}
 -\frac{\sqrt{2} A_{D^0\to \phi_1  K^0}}{\lambda^2}& -\frac{A_{D^0\to \phi_1  \pi_0}}{ \lambda }+\frac{\sqrt{3}A_{D^0\to \phi_1  \eta_8}}{\lambda }+\sqrt{2} A_{D^0\to \phi_1   \bar{K}^0} =0 ~.
\end{align*}
\eenum

\subsection{Isospin Sum Rules}
As for the $PP$ case, we now proceed to determine the $P_8V_8$ isospin sum rules. Similarly to eq. (\ref{eqn:DPII}) the isospin $\times$ strangeness vector meson irreps are
\begin{gather}
	\rho^i_j = \begin{pmatrix} \frac{\rho^0}{\sqrt{2}} & \rho^+ \\ \rho^- & -\frac{\rho^0}{\sqrt{2}} \end{pmatrix} \sim 3_0~, \quad K^{*i} = \begin{pmatrix} K^{*+} \\ K^{*0} \end{pmatrix} \sim 2_{-1}~,\quad \omega_{8},\phi_{1} \sim 1_0~,\notag\\
	 \bar{K}^*_i = \begin{pmatrix} K^{*-} & \bar{K}^{*0} \end{pmatrix} \sim 2_{+1}~.
\end{gather}
This time there are 6 isospin sum rules; two for the $\Pi \rho$ final state, and one each for the $\rho K$, $\rho \bar{K}$, $\Pi K^*$ and $\Pi \bar{K}^*$ final states. Explicitly, these are
\benum
\item \label{sum:PVI1}
\begin{align*}
 A_{D^+_s\to \rho _0 \pi ^+}& +A_{D^+_s\to \pi _0 \rho ^+} =0
\end{align*}
\item \label{sum:PVI2}
\begin{align*}
 \sqrt{2} A_{D^0\to \rho _0 \bar{K}^0}&+A_{D^0\to K^- \rho ^+}- A_{D^+\to \bar{K}^0 \rho ^+} =0
\end{align*}
\item \label{sum:PVI3}
\begin{align*}
 \sqrt{2}A_{D^0\to \pi _0\bar{K}^{*0}}&+ A_{D^0\to K^{*-} \pi ^+}-A_{D^+\to\bar{K}^{*0} \pi ^+} =0
\end{align*}
\item \label{sum:PVI4}
\begin{align*}
 A_{D^0\to \pi _0 K^{*0}}&+\frac{A_{D^0\to \pi ^-K^{*+}}}{\sqrt{2}}-\frac{A_{D^+\to K^{*0} \pi ^+}}{\sqrt{2}}+A_{D^+\to \pi _0K^{*+}}
   =0
\end{align*}
\item \label{sum:PVI5}
\begin{align*}
 A_{D^0\to K^0 \rho _0}&+\frac{A_{D^0\to \rho ^- K^+}}{\sqrt{2}}+A_{D^+\to \rho _0 K^+}-\frac{A_{D^+\to K^0 \rho ^+}}{\sqrt{2}} =0
\end{align*}
\item \label{sum:PVI6}
\begin{align*}
 -\sqrt{2} A_{D^0\to \pi _0 \rho _0}&+\frac{A_{D^0\to \rho ^- \pi ^+}}{\sqrt{2}}+\frac{A_{D^0\to \pi ^- \rho ^+}}{\sqrt{2}}+A_{D^+\to \rho _0 \pi ^+}+A_{D^+\to
   \pi _0 \rho ^+} =0~.
\end{align*}
\eenum
Again, these isospin sum rules hold to all s-mass spurion orders. Computing isospin breaking invariants  one finds at first order in the isospin breaking spurion that $\mbox{\ref{sum:PVI2}} + \mbox{\ref{sum:PVI3}} -\sqrt{2}\mbox{\ref{sum:PVI1}}$ is valid to $\mathcal{O}(\delta^2)$, as are sum rules \ref{sum:PVI4}, \ref{sum:PVI5} and \ref{sum:PVI6}. These four sum rules thus provide further isospin tests of the pattern of SU(3) breaking, that are highly sensitive to new physics. Perhaps the easiest to measure is the Cabibbo-favored combination $\mbox{\ref{sum:PVI2}} + \mbox{\ref{sum:PVI3}} -\sqrt{2}\mbox{\ref{sum:PVI1}}$, which is equivalent to measuring the deviation from zero of the reduced amplitude relation
\begin{equation}
	\label{eqn:PVTB}
	\frac{ \sqrt{2} \amB{D^0}{ \rho _0 \bar{K}^0}+\amB{D^0}{ K^- \rho ^+} + \sqrt{2}\amB{D^0}{ \pi _0\bar{K}^{*0}}+ \amB{D^0}{ K^{*-} \pi ^+}}{\amB{D^+}{ \bar{K}^0 \rho ^+} + \amB{D^+}{\bar{K}^{*0} \pi ^+} + \sqrt{2}\amB{D^+_s}{ \rho _0 \pi ^+} +\sqrt{2}\amB{D^+_s}{ \pi _0 \rho ^+}} -1~.
\end{equation}
At present, not all these modes have been measured \cite{pdg:2012}, and moreover, one must find the strong phases.

\subsection{\texorpdfstring{$P_8 V_8$}{OE}  Sum Rules}
\label{sec:PV88SR}
Returning to SU(3) , the two U-spin sum rules are
\benum
\item \label{sum:PVUA1}
\begin{align*}
 -A_{D^0\to K^{*-} \pi ^+} &+\frac{A_{D^0\to K^{*-} K^+} }{\lambda} -\frac{A_{D^0\to \rho ^- \pi ^+} }{\lambda} +\frac{A_{D^0\to \rho ^- K^+}}{\lambda^2} =0
\end{align*}
\item \label{sum:PVUA2}
\begin{align*}
 A_{D^0\to K^- \rho ^+}&+\frac{A_{D^0\to \pi ^- \rho ^+}}{\lambda }-\frac{A_{D^0\to K^-K^{*+}}}{\lambda }-\frac{A_{D^0\to \pi ^- K^{*+}}}{\lambda ^2} =0~.
\end{align*}
\eenum
Choosing a basis in which the isospin and U-spin sum rules are explicit, we find a further 15 pure SU(3) sum rules, valid to $\mathcal{O}(\varepsilon^2)$, 
\benum
\item
\begin{align*}
 \frac{A_{D^+\to\bar{K}^{*0} K^+}}{\lambda}&-\frac{A_{D^+\to K^{*0} \pi ^+}}{\lambda^2 }-  A_{D^+\to\bar{K}^{*0} \pi ^+}+\frac{A_{D^+_s\to K^{*0}
   K^+}}{\lambda^2 }\\[10pt]
   &+  A_{D^+_s\to\bar{K}^{*0} K^+}-\frac{A_{D^+_s\to K^{*0} \pi ^+}}{\lambda} =0
\end{align*}
\item
\begin{align*}
 A_{D^+\to \bar{K}^0 \rho ^+}&-A_{D^+_s\to \bar{K}^0K^{*+}}-\frac{A_{D^+\to \bar{K}^0K^{*+}} }{\lambda}
   +\frac{A_{D^+_s\to K^0 \rho ^+}}{ \lambda}\\[10pt]
   & +\frac{A_{D^+\to K^0 \rho ^+}}{\lambda^2}-\frac{A_{D^+_s\to K^0K^{*+}}}{\lambda^2} =0
\end{align*}
\item
\begin{align*}
\sqrt{\frac{3}{2}}  \frac{A_{D^0\to \eta_8  K^{*0}}}{\lambda^2 }&-\frac{A_{D^0\to \pi _0 K^{*0}}}{\sqrt{2} \lambda^2 }+\frac{A_{D^0\to \bar{K}^0
   K^{*0}}}{\lambda}+\sqrt{\frac{3}{2}}   \frac{A_{D^0\to \eta_8 \bar{K}^{*0}}}{\lambda}\\[10pt]
   &-\frac{A_{D^0\to K^0\bar{K}^{*0}}}{\lambda^2}-\frac{ A_{D^0\to \pi _0\bar{K}^{*0}}}{\sqrt{2}\lambda} =0
\end{align*}
\item
\begin{align*}
 -\frac{A_{D^0\to \eta_8  \omega_8 }}{\lambda}&+\sqrt{\frac{2}{3}}\frac{ A_{D^0\to \omega_8  K^0}}{\lambda^2 }+\frac{A_{D^0\to \omega_8  \pi _0}}{\sqrt{3}\lambda}-\sqrt{\frac{2}{3}} 
   A_{D^0\to \omega_8  \bar{K}^0}+\sqrt{\frac{3}{2}} \frac{A_{D^0\to \eta_8  K^{*0}}}{\lambda^2 }\\[10pt]
   &-\frac{A_{D^0\to \pi _0 K^{*0}}}{\sqrt{2} \lambda^2 }+\frac{A_{D^0\to\bar{K}^0 K^{*0}}}{\lambda} =0
\end{align*}
\item
\begin{align*}
 -\frac{A_{D^0\to \eta_8  \omega_8 }}{\lambda}&+\sqrt{\frac{3}{2}}\frac{ A_{D^0\to \omega_8  K^0}}{\lambda^2 }+\frac{A_{D^0\to \eta_8  \rho _0}}{\sqrt{3}\lambda}-\frac{A_{D^0\to K^0 \rho
   _0}}{\sqrt{2} \lambda ^2}+\sqrt{\frac{2}{3}} \frac{A_{D^0\to \eta_8  K^{*0}}}{\lambda^2 }\\[10pt]
   &-\sqrt{\frac{2}{3}}   A_{D^0\to \eta_8 \bar{K}^{*0}}+\frac{A_{D^0\to K^0 \bar{K}^{*0}} }{\lambda}=0
\end{align*}
\item
\begin{align*}
 \frac{A_{D^+_s\to \omega_8  K^+}}{\lambda}&-\frac{A_{D^+_s\to \rho _0 K^+}}{\sqrt{3}\lambda}-\sqrt{\frac{2}{3}} \frac{A_{D^+_s\to K^{*0}
   K^+}}{\lambda^2 }+\sqrt{\frac{2}{3}} A_{D^+_s\to\bar{K}^{*0} K^+}-A_{D^+_s\to \omega_8  \pi ^+}\\[10pt]
   &+\frac{A_{D^+_s\to \rho _0 \pi ^+}}{\sqrt{3}}+\sqrt{\frac{2}{3}} \frac{A_{D^+_s\to K^{*0} \pi ^+}}{\lambda} =0
\end{align*}
\item
\begin{align*}
 \frac{A_{D^+_s\to \rho _0 \pi ^+}}{\sqrt{3}}&+A_{D^+_s\to \eta_8  \rho ^+}-\sqrt{\frac{2}{3}}\frac{ A_{D^+_s\to K^0
   \rho ^+}}{\lambda }-\frac{A_{D^+_s\to \eta_8 K^{*+}}}{\lambda }+\sqrt{\frac{2}{3}}\frac{ A_{D^+_s\to K^0
  K^{*+}}}{\lambda ^2}\\[10pt]
   &+\frac{A_{D^+_s\to \pi _0K^{*+}}}{\sqrt{3} \lambda }-\sqrt{\frac{2}{3}}A_{D^+_s\to \bar{K}^0K^{*+}} =0
\end{align*}
\item
\begin{align*}
 \frac{A_{D^+\to \omega_8  \pi ^+}}{\lambda}&-\frac{A_{D^+\to \rho _0 \pi ^+}}{\sqrt{3}\lambda}-\sqrt{\frac{2}{3}} \frac{A_{D^+\to K^{*0} \pi ^+}}{\lambda^2 }+\sqrt{\frac{2}{3}} 
   A_{D^+\to\bar{K}^{*0} \pi ^+}+  A_{D^+_s\to \omega_8  \pi ^+}\\[10pt]
   &-\frac{  A_{D^+_s\to \rho _0 \pi
   ^+}}{\sqrt{3}}-\sqrt{\frac{2}{3}} \frac{A_{D^+_s\to K^{*0} \pi ^+}}{\lambda} =0
\end{align*}
\item
\begin{align*}
 -A_{D^+_s\to \rho _0 \pi ^+}&-\frac{A_{D^+\to \rho _0 \pi ^+}}{ \lambda }+\frac{A_{D^+_s\to \rho _0 K^+}}{ \lambda}
   -\frac{A_{D^+_s\to K^{*0} \pi ^+} }{\sqrt{2}\lambda}+\frac{A_{D^+\to \rho _0 K^+}}{\lambda^2}\\[10pt]
   &-\frac{A_{D^+\to K^{*0} \pi^+}}{\sqrt{2}\lambda^2}+\frac{A_{D^+_s\to K^{*0} K^+}}{\sqrt{2}\lambda^2} =0
\end{align*}
\item
\begin{align*}
 \frac{A_{D^+\to \pi _0 \rho ^+}}{\lambda}&-\frac{ A_{D^+\to \bar{K}^0 \rho ^+}}{\sqrt{2}}-\frac{A_{D^+\to \pi _0K^{*+}}}{\lambda^2 }+\frac{A_{D^+\to
   \bar{K}^0K^{*+}}}{\sqrt{2}\lambda}- A_{D^+_s\to \rho _0 \pi ^+}\\[10pt]
   &-\frac{A_{D^+_s\to \pi _0 K^{*+}}}{\lambda}+\frac{  A_{D^+_s\to \bar{K}^0K^{*+}}}{\sqrt{2}} =0
\end{align*}
\item
\begin{align*}
 \sqrt{\frac{2}{3}} A_{D^+_s\to \bar{K}^0K^{*+}}&+\sqrt{\frac{2}{3}} \frac{A_{D^+\to \bar{K}^0K^{*+}}}{ \lambda}
   +\frac{A_{D^+_s\to \eta_8 K^{*+}} }{\lambda }-\frac{A_{D^+_s\to \pi _0K^{*+}} }{\sqrt{3}\lambda}+\frac{A_{D^+\to \eta_8 K^{*+}}}{\lambda^2}\\[10pt]
   &-\frac{A_{D^+\to \pi _0K^{*+}}}{\sqrt{3}\lambda^2}-\sqrt{\frac{2}{3}} \frac{A_{D^+_s\to K^0K^{*+}} }{\lambda^2}=0
\end{align*}
\item
\begin{align*}
 \sqrt{\frac{2}{3}} A_{D^+_s\to \eta_8  \rho ^+}&+\sqrt{\frac{2}{3}} \frac{A_{D^+\to \eta_8  \rho ^+}}{ \lambda} -\frac{A_{D^+_s\to K^0
   \rho ^+} }{\lambda} -\sqrt{\frac{2}{3}} \frac{A_{D^+_s\to \eta_8 K^{*+}} }{\lambda} -\frac{A_{D^+\to K^0 \rho ^+}}{\lambda^2}\\[10pt]
   &-\sqrt{\frac{2}{3}} \frac{A_{D^+\to \eta_8  K^{*+}}}{\lambda^2}+\frac{A_{D^+_s\to K^0K^{*+}}}{\lambda^2} =0
\end{align*}
\item
\begin{align*}
 -2 A_{D^0\to \pi _0\bar{K}^{*0}}&-\sqrt{6} \frac{A_{D^0\to \omega_8  \pi _0} }{\lambda} +\sqrt{2} \frac{A_{D^0\to \pi _0 \rho _0}}{ \lambda} -\sqrt{2} \frac{A_{D^0\to K^0
  \bar{K}^{*0}}}{ \lambda} -\sqrt{3} \frac{A_{D^0\to \omega_8  K^0}}{\lambda^2}\\[10pt]
   &+\frac{A_{D^0\to K^0 \rho _0}}{\lambda^2}+2 \frac{A_{D^0\to \pi _0 K^{*0}}}{\lambda^2} =0
\end{align*}
\item
\begin{align*}
 \frac{\sqrt{\frac{3}{2}} A_{D^0\to \eta_8  \rho _0}}{\lambda }&-\frac{A_{D^0\to K^0 \rho _0}}{\lambda ^2}-\frac{A_{D^0\to \pi _0 \rho _0}}{\sqrt{2} \lambda
   }+A_{D^0\to \rho _0 \bar{K}^0}-\frac{1}{2} \sqrt{3} A_{D^0\to \eta_8 \bar{K}^{*0}}\\[10pt]
   &+\frac{A_{D^0\to K^0\bar{K}^{*0}}}{\sqrt{2} \lambda }+\frac{1}{2}
   A_{D^0\to \pi _0\bar{K}^{*0}} =0
\end{align*}
\item
\begin{align*}
 \sqrt{\frac{2}{3}} A_{D^+\to\bar{K}^{*0} \pi ^+}&-\sqrt{\frac{2}{3}} A_{D^+_s\to\bar{K}^{*0} K^+}+A_{D^+_s\to \omega_8  \pi ^+} -\frac{2 A_{D^+_s\to \rho _0 \pi ^+}}{\sqrt{3}}-\frac{A_{D^+\to \rho _0 \pi^+} }{\sqrt{3}\lambda}\\[10pt]
   +\frac{A_{D^+_s\to \rho _0 K^+}  }{\sqrt{3}\lambda}&-\frac{A_{D^+_s\to K^{*0} \pi ^+}
   }{\sqrt{6}\lambda}+\frac{A_{D^+\to \omega_8  K^+}}{\lambda^2}+\frac{A_{D^+\to K^{*0} \pi ^+}}{\sqrt{6}\lambda^2}-\frac{A_{D^+_s\to K^{*0} K^+}}{\sqrt{6}\lambda^2}
=0~.
\end{align*}
\eenum
Rotation to the mass basis follows from Sec. \ref{sec:FM}, noting that in the $D \to PV$ case there are no amplitudes requiring the symmetry factor of eq. (\ref{eqn:AMB}). Unlike in the $PP$ case, one may also measure modes involving $K^{*0}$ and $\bar{K}^{*0}$ directly, via tagging with $K$ or $K_S$, so that we need not rotate to $K^*$ mass basis. 

\subsection{Rate Sum Rules}
\label{sec:PVRSR}
We finally present the $D \to PV$ rate sum rules valid to $\mathcal{O}(\varepsilon^2)$, in the mass eigenstate basis, of which there are six:

\setcounter{incre}{0}

\benum
\item
\begin{gather*}
 |\amB{D_s^+}{\rho_0\pi^+}|^2=|\amB{D_s^+}{\pi_0 \rho^+}|^2 
\end{gather*}
\item \label{sum:PVU1}
\begin{gather*}
 \frac{|\amB{D^0}{\rho^- K^+}|^2}{\lambda^4}+|\amB{D^0}{K^{*-} \pi^+}|^2 = \frac{|\amB{D^0}{K^{*-} K^+}|^2}{\lambda^2}+\frac{|\amB{D^0}{\rho^- \pi^+}|^2}{\lambda^2}
\end{gather*}
\item \label{sum:PVU2}
\begin{gather*}
 |\amB{D^0}{K^- \rho^+}|^2+\frac{|\amB{D^0}{\pi^- K^{*+}}|^2}{\lambda^4} = \frac{|\amB{D^0}{\pi^- \rho^+}|^2}{\lambda^2}+\frac{|\amB{D^0}{K^- K^{*+}}|^2}{\lambda^2}
\end{gather*}
\item
\begin{gather*}
   \frac{|\amB{D^+}{\eta  \rho^+}|^2}{\lambda^2}+\bigg(\frac{|\amB{D^+}{K_S \rho^+}|^2}{\lambda^2}- \frac{|\amB{D^+}{K_L \rho^+}|^2}{\lambda^2}\bigg)+\frac{|\amB{D^+}{\pi_0 \rho^+}|^2}{\lambda^2}\\
  +\frac{|\amB{D^+}{\rho^+ \eta '}|^2}{\lambda^2} +\bigg[\frac{|\amB{D_s^+}{K_S K^{*+}}|^2}{\lambda^2}-\frac{|\amB{D_s^+}{K_L K^{*+}}|^2}{\lambda^2}\bigg] +\frac{|\amB{D_s^+}{\eta  K^{*+}}|^2}{\lambda^2}\\
   +\frac{|\amB{D_s^+}{\pi_0K^{*+}}|^2}{\lambda^2}+\frac{|\amB{D_s^+}{K^{*+} \eta '}|^2}{\lambda^2} \\
  =   \\
  \frac{|\amB{D^+}{\eta  K^{*+}}|^2}{\lambda^4}+\frac{|\amB{D^+}{\pi_0 K^{*+}}|^2}{\lambda^4}+\frac{|\amB{D^+}{K^{*+} \eta '}|^2}{\lambda^4}+|\amB{D_s^+}{\eta  \rho^+}|^2\\
 +|\amB{D_s^+}{\pi_0 \rho^+}|^2+|\amB{D_s^+}{\rho^+ \eta '}|^2
\end{gather*}
\item
\begin{gather*}
  \frac{|\amB{D^+}{\bar{K}^{*0} K^+}|^2}{\lambda^2}+\frac{|\amB{D^+}{\phi  \pi^+}|^2}{\lambda^2}+\frac{|\amB{D^+}{\omega  \pi^+}|^2}{\lambda^2}+\frac{|\amB{D^+}{\rho_0\pi^+}|^2}{\lambda^2} +\frac{|\amB{D_s^+}{\phi  K^+}|^2}{\lambda^2}\\
  +\frac{|\amB{D_s^+}{\omega K^+}|^2}{\lambda^2}+\frac{|\amB{D_s^+}{\rho_0 K^+}|^2}{\lambda^2}+\frac{|\amB{D_s^+}{K^{*0} \pi^+}|^2}{\lambda ^2} \\
  =\\
   \frac{|\amB{D^+}{\phi  K^+}|^2}{\lambda^4}+\frac{|\amB{D^+}{\omega  K^+}|^2}{\lambda^4}+\frac{|\amB{D^+}{\rho_0 K^+}|^2}{\lambda^4}+\frac{|\amB{D^+}{K^{*0} \pi^+}|^2}{\lambda^4}+|\amB{D^+}{\bar{K}^{*0} \pi^+}|^2\\
   +\frac{|\amB{D_s^+}{K^{*0}K^+}|^2}{\lambda^4}  +|\amB{D_s^+}{\bar{K}^{*0} K^+}|^2+|\amB{D_s^+}{\phi  \pi^+}|^2+|\amB{D_s^+}{\omega  \pi^+}|^2+|\amB{D_s^+}{\pi_0 \rho^+}|^2
\end{gather*}
\item
\begin{gather*}
\bigg[\frac{|\amB{D^0}{\phi  K_L}|^2}{\lambda^2} - \frac{|\amB{D^0}{\phi  K_S}|^2}{\lambda^2}\bigg]+\bigg[\frac{|\amB{D^0}{\omega  K_L}|^2}{\lambda^2} - \frac{|\amB{D^0}{\omega  K_S}|^2}{\lambda ^2}\bigg]\\
+\bigg[\frac{|\amB{D^0}{K_L \rho_0}|^2}{\lambda^2}- \frac{|\amB{D^0}{K_S \rho_0}|^2}{\lambda^2}\bigg]+\frac{|\amB{D^0}{\eta  K^{*0}}|^2}{\lambda^4}+\frac{|\amB{D^0}{\pi_0 K^{*0}}|^2}{\lambda^4} +\frac{|\amB{D^0}{K^{*0}\eta '}|^2}{\lambda^4} \\
+|\amB{D^0}{\eta  \bar{K}^{*0}}|^2+|\amB{D^0}{\pi_0 \bar{K}^{*0}}|^2+|\amB{D^0}{\bar{K}^{*0} \eta '}|^2\\
  = \\
  \frac{|\amB{D^0}{\eta  \phi }|^2}{\lambda^2}+\frac{|\amB{D^0}{\eta \omega }|^2}{\lambda^2}+\frac{|\amB{D^0}{\phi  \pi_0}|^2}{\lambda^2}+\frac{|\amB{D^0}{\omega  \pi_0}|^2}{\lambda^2}+\frac{|\amB{D^0}{\eta  \rho_0}|^2}{\lambda^2} +\frac{|\amB{D^0}{\pi_0 \rho_0}|^2}{\lambda^2}\\
  +\frac{|\amB{D^0}{\phi  \eta '}|^2}{\lambda^2}+\frac{|\amB{D^0}{\omega  \eta '}|^2}{\lambda^2}+\frac{|\amB{D^0}{\rho_0 \eta '}|^2}{\lambda^2}~,
\end{gather*}
\eenum
and a further four sum rules that result just from the $K$-$K$ mixing,
\benum
\item
\begin{gather*}
 |\amB{D^0}{K_S K^{*0}}|^2=|\amB{D^0}{K_L K^{*0}}|^2 
\end{gather*}
\item
\begin{gather*}
 |\amB{D^0}{K_S \bar{K}^{*0}}|^2=|\amB{D^0}{K_L  \bar{K}^{*0}}|^2 
\end{gather*}
\item
\begin{gather*}
 |\amB{D^+}{K_S K^{*+}}|^2=|\amB{D^+}{K_L K^{*+}}|^2 
\end{gather*}
\item
\begin{gather*}
   |\amB{D_s^+}{K_S \rho^+}|^2=|\amB{D_s^+}{K_L \rho^+}|^2 ~.
\end{gather*}
\eenum

\subsection{PV Predictions}
The rate sum rules \ref{sum:PVU1} and \ref{sum:PVU2} in Sec. \ref{sec:PVRSR} are the PV equivalent of the PP U-spin rate sum rule. Combining these respectively with the U-spin amplitude sum rules \ref{sum:PVUA1} and \ref{sum:PVUA2} of Sec. \ref{sec:PV88SR}, one may also show (see Sec. \ref{sec:PPRSR}) that the following normed amplitude sum rules also hold to $\mathcal{O}(\varepsilon^2)$:
\begin{gather}
	|\amB{D^0}{\pi^+ \rho^-}|/\lambda +  |\amB{D^0}{K^+ K^{*-}}|/\lambda = |\amB{D^0}{K^+ \rho^-}|/\lambda^2 + |\amB{D^0}{\pi^+ K^{*-}}| \label{eqn:PVNU1} \\
	|\amB{D^0}{\pi^- \rho^+}|/\lambda +  |\amB{D^0}{K^- K^{*+}}|/\lambda = |\amB{D^0}{K^- \rho^+}| + |\amB{D^0}{\pi^- K^{*+}}|/\lambda^2~. \label{eqn:PVNU2}
\end{gather}
We again emphasize that these normed amplitude sum rules are a consequence of the fact that there are both amplitude and rate sum rules involving the same modes, and this circumstance unique to the U-spin sum rules.

The branching ratios of \ref{eqn:PVNU2} have been measured, and one finds from the data\footnote{Particular care must be taken with the current PDG data for modes with $K^{*\pm}$ in the final state. At present, the  $D^0 \to K^{*+} K^-$ mode has only been measured for the case that $K^{*+}$ subsequently decays to $K^+\pi^0$. The isospin Clebsch-Gordan coefficients imply this occurs one-third of the time while $K^{*+} \to K^0\pi^+$ occurs two-thirds of the time. As a result, we must multiply the current PDG rate for $D^0 \to (K^{*+}) K^- \to (K^+\pi^0)K^-$ by this factor of three. Similar care must be taken with the data for  $D^0 \to K^{*+} \pi^-$ and $D^0 \to K^{*-} K^+$. Both $K^{*-} \to K^-\pi^0$ and $K^{*-} \to K_S\pi^+$ decay channels have been measured for the $D^0 \to K^{*-} \pi^+$ mode, so in this case we na\" \i vely average the rates with appropriate Clebsch-Gordan factors.}  \cite{pdg:2012} 
\begin{equation}
	\label{eqn:PVU1}
	\frac{|\amB{D^0}{\pi^- \rho^+}|/\lambda+ |\amB{D^0}{K^- K^{*+}}|/\lambda}{ |\amB{D^0}{K^- \rho^+}|+|\amB{D^0}{\pi^- K^{*+}}|/\lambda^2} -1 = 6\% \pm 17\%~,
\end{equation}
which is comparable to the PP U-spin sum rule (\ref{eqn:USSR}), though less precise. The DCS process of \ref{eqn:PVNU1} is yet to be measured, so from the corresponding rate sum rule we instead obtain the prediction \cite{pdg:2012}
\begin{equation}
	\label{eqn:PVU2}
	\mbox{Br}(D^0 \to \rho^- K^+) \simeq (1.7 \pm 0.4) \times 10^{-4}~.
\end{equation}

We note further that we have from the data \cite{pdg:2012}
\begin{equation}
	\label{eqn:PVUB}
	\bigg|\frac{\amB{D^0}{\pi^+\rho^-}}{\amB{D^0}{K^+K^{*-}}}\bigg| - 1= 0.59 \pm 0.10~, \qquad \bigg|\frac{\amB{D^0}{\pi^-\rho^+}}{\amB{D^0}{K^-K^{*+}}}\bigg| - 1 = 0.33 \pm 0.05~.
\end{equation}
Compared to the PP case in eq. (\ref{eqn:KPR}), this implies a slightly smaller and inverse U-spin breaking for PV. If the prediction (\ref{eqn:PVU2}) is satisfied, then eqs. (\ref{eqn:PVU1}) -- (\ref{eqn:PVUB}) are consistent with the $\Delta U = 0$ rule proposed in Ref. \cite{Brod:2012ud} for U-spin irreps, just as eqs. (\ref{eqn:KPR}) and (\ref{eqn:USSR}) are consistent with this rule for the PP case.  To be explicit, under the $\Delta U = 0$ rule one assumes a large broken penguin picture, in which
\begin{gather}
	\amB{D^0 }{ K^\pm K^{*\mp}}  \simeq \lambda [T^\pm -P_{\rm b}^\pm ]- \lambda^5e^{i(\delta^\pm - \gamma)} P^\pm ~,\quad \amB{D^0 }{ \pi^\pm \rho^\mp}  \simeq -\lambda [T^\pm + P_{\rm b}^\pm] - \lambda^5e^{i(\delta^\pm - \gamma)} P^\pm~, \notag\\
	\amB{D^0}{ K^{*+}\pi^-}  \simeq \lambda^2T^+~,\quad  \amB{D^0}{ K^{*-}\pi^+} \simeq T^- ~,\quad \amB{D^0}{\rho^+ K^{-}}  \simeq T^+~,\quad \amB{D^0}{\rho^- K^{+}} \simeq \lambda^2T^-~, \label{eqn:BPP}
\end{gather}
where $T$ and $P$ are respectively U-spin tree and penguin reduced matrix elements, $P_b$ is the so-called broken penguin --  which is a U-spin breaking reduced matrix element, na\"\i vely $\mathcal{O}(\varepsilon)$ -- and $\delta^\pm$ are ($\gamma$ is) the strong phases (weak phase). It is assumed that the penguins are enhanced, such that $P \sim \mathcal{O}(1/\varepsilon)$ and $P_{\rm b} \sim \mathcal{O}(1)$, while $T$ remains $\mathcal{O}(1)$.  Eqs. (\ref{eqn:BPP}) are consistent both with the PV U-spin sum rules as well as $\mathcal{O}(1)$ breakings of eqs. (\ref{eqn:PVUB}).  

Applying this U-spin picture, one may predict the ratio of the $K K^*$ and $\pi \rho$ direct CP asymmetries,
\begin{equation}
	\frac{\Acp(K^{\pm} K^{*\mp})}{\Acp(\pi^{\pm}\rho^{\mp})}  \simeq \frac{\amB{D^0}{\pi^\pm \rho^\mp}}{\amB{D^0}{K^\pm K^{*\mp}}} + \mathcal{O}(\varepsilon)~.
\end{equation}
We expect $\amB{D^0}{K^\pm \rho^\mp}$ to be opposite sign to $\amB{D^0}{\pi^\pm \rho^\mp}$ in the $\varepsilon \to 0$ limit and at leading order in $\lambda$  (see Appendix \ref{app:PV} and eq. (\ref{eqn:BPP})), so from eqs. (\ref{eqn:PVUB}) we have
\begin{equation}
\label{eqn:ACPR}
	\frac{\Acp(K^{+} K^{*-})}{\Acp(\pi^{+}\rho^{-})} \simeq -1.59 \pm 0.10~, \quad \mbox{and} \quad \frac{\Acp(K^{-} K^{*+})}{\Acp(\pi^{-}\rho^{+})} \simeq -1.33 \pm 0.05~,
\end{equation}
up to $\mathcal{O}(\varepsilon)$ corrections. 

We may further estimate the PV $\Delta\Acp$'s, defined to be
\begin{align}
	\Delta \Acp^\pm 
	& \equiv \Acp(D^0 \to K^\pm K^{*\mp}) - \Acp(D^0 \to \pi^\pm \rho^\mp)\notag\\
	& = -2\lambda^4\bigg(\frac{P^\pm}{T^\pm - P_{\rm b}^\pm} + \frac{P^\pm}{T^\pm + P_{\rm b}^\pm}\bigg)\sin\delta\sin\gamma~.
\end{align}
Let $P^0$, $P^0_{\rm b}$, $T^0$ and $\delta_0$, be the penguin, broken penguin and tree terms and strong phase respectively of the PP system, and define $\mathcal{R}^\pm \equiv P_{\rm b}^\pm/T^\pm$ and $\mathcal{R}^0 \equiv P^0_{\rm b}/T^0$. Assuming that PP and PV have same penguin contraction enhancements, such that $P^0/P^0_{\rm b} \simeq P^\pm/P^\pm_{\rm b}$, then it follows that 
\begin{equation}
	\Delta \Acp^\pm \simeq \Delta \Acp \bigg[\frac{\sin \delta^\pm}{\sin \delta_0}\bigg]\bigg[ \frac{(\mathcal{R}^{\pm})^2}{1 - (\mathcal{R}^{\pm})^2}\bigg] \bigg[\frac{1 -(\mathcal{R}^{0})^2}{(\mathcal{R}^{0})^2}\bigg]~.
\end{equation}
Since $PV $ is a spin-1 final state, we expect $\sin \delta^\pm/\sin \delta_0 \sim -1$ and from the data \cite{pdg:2012}
\begin{equation}
	\label{eqn:RV}
	\mathcal{R}^+ = 0.23 \pm 0.03~,\qquad \mathcal{R}^-= 0.14 \pm 0.02~, \qquad  \mathcal{R}^0 = -0.29 \pm 0.01~.
\end{equation}
We then estimate up to $\mathcal{O}(1)$ uncertainty
\begin{equation}
	 \label{eqn:ACPP}
	\Delta \Acp^\pm \sim -\Delta \Acp~.
\end{equation}
We emphasize that these predictions pertain only to $\Delta U = 0$ rule of Ref.  \cite{Brod:2012ud}, and they are independent from the SU(3) sum rule analysis of this paper. Nonetheless, the analysis of this section is motivated by our prediction that the U-spin sum rules (\ref{eqn:PVNU1}) and (\ref{eqn:PVNU2}) are valid to $\mathcal{O}(\varepsilon^2)$ -- and one of them appears to be valid to this order --  and this prediction is consistent with the $\Delta U = 0$ rule for large broken penguin picture.

\section{Summary}
In this paper we have presented the amplitude and rate sum rules, valid to $\mathcal{O}(\varepsilon^2)$,  associated with SU(3) breaking by the $m_s$ spurion for both $D \to PP$ and $D \to PV$ decays. At the amplitude level, verifying these sum rules provides a test of this pattern of flavor SU(3) breaking. In particular, the isospin sum rules (\ref{eqn:PTB}) and (\ref{eqn:PVTB}) that are valid to second order in $m_I$, provide an extremely sensitive test of new SU(3) breaking sources. 

In practical terms, testing the amplitude level sum rules will prove difficult in the immediate future, because of the need to measure the strong phases. As a result, the square ampltiude or rate sum rules are a better candidate for future experimental tests, in particular the  $PV$ U-spin sum rules (\ref{eqn:PVU1}) and (\ref{eqn:PVU2}). The so-far imprecise verification of (\ref{eqn:PVU1}) is encouraging for the development of a large broken penguin $\Delta U =0 $ rule for the PV case, analogous to Ref. \cite{Brod:2012ud}. Such a rule implies the predictions for the PV direct CP asymmetries (\ref{eqn:ACPR}) and (\ref{eqn:ACPP}) that we have provided above.

Finally, let us briefly discuss the applicability of our $D\to PP$ results to $D \to VV$, which we have not considered explicitly in this paper. For $D \to VV$, the extra Lorentz structure of the meson tensors -- that is we have $V^\mu V^\nu$ -- means that the final states can be CP even or CP odd, compared to PP in which all final states are CP even. This yields a larger number of invariants and corresponding reduced matrix elements. Put in other words, whereas in PP the symmetry of the final states restricts us to symmetrized tensor contractions, in VV there is no such restriction. Despite this complication, the small phase space available to most VV decays implies that they are dominated by the s-wave channel. As a result, simply replacing P mesons with V mesons everywhere in the PP results will provide approximately correct $D \to VV$ relations.

\acknowledgments{ The authors thank J. Zupan for helpful discussions. This work is supported by the U.S. National Science Foundation through grant PHY-0757868 and by the United States-Israel Binational
Science Foundation (BSF) under grant No. 2010221.}

\appendix

\section{Abstract Sum Rule Generation}
\label{app:ASRG}
In this appendix, we show how to compute sum rules from the symmetries of the Hamiltonian $H$. Examining the structure of $H$ in eqs. (\ref{eqn:CKMH}), note first that the $\bm{6}$ can be written in matrix form as $[\bm{6}]^i_j \equiv [\bm{6}]^{ij}$, i.e (dropping factors of 1/2 and overall signs)
\begin{equation}
	[\bm{6}] = \begin{pmatrix} 0 & 0 &0 \\ 0 & 1 & \lambda \\ 0 & \lambda & \lambda^2 \end{pmatrix}~,
\end{equation}
with tensor transformation law under a generator $X$  
\begin{equation}
	[\bm{6}]^i_j \to \{X[\bm{6}]\}^i_j + \{X[\bm{6}]^T\}^i_j~.
\end{equation}
This is clearly zero for $X = E_\pm^I$ and $E_\pm^D$, the raising/lowering operators (normalized to unity) of isospin and D-spin respectively. Similarly, defining the matrices $[\bm{\bar{15}}]_i$ via
\begin{equation}
	([\bm{\bar{15}}]_i)^k_j \equiv [\bm{\bar{15}}]_{ij}^k~,
\end{equation}
then in matrix notation we may write the $\bm{\bar{15}}$ irrep as
\begin{equation}
	[\bm{\bar{15}}]_1 =\begin{pmatrix}0 & 0 & 0\\ 0& -\lambda & 1\\0 &-\lambda^2 & \lambda \end{pmatrix}
	\qquad [\bm{\bar{15}}]_2 =\begin{pmatrix} 0 & 0 & 0\\ -\lambda&0&0 \\ -\lambda^2&0&0 \end{pmatrix}
	\qquad [\bm{\bar{15}}]_3 =\begin{pmatrix}0&0&0 \\1 &0& 0\\\lambda &0& 0 \end{pmatrix}~,
\end{equation}
and by symmetry of the lower indices, the tensor transformation law under generator $X$ is
\begin{equation}
	 [\bm{\bar{15}}]_{ij}^k \to 2 \{[\bm{\bar{15}}]^{\phantom{k}}_{(i}X\}^k_{j)} -  \{X[\bm{\bar{15}}]_i\}^k_j~.
\end{equation}
This is zero under the operator
\begin{equation}
	T_- \equiv E_-^I + \lambda E_-^D = \begin{pmatrix}0&0&0 \\1 &0& 0\\\lambda &0& 0 \end{pmatrix}~.
\end{equation}
Clearly both $\bm{\bar{3}_p}$ and $\bm{\bar{3}_t}$ have matrix form $\sim(\lambda^5A^2,0,0)$, so that they are invariant under this operator too. Consequently, we deduce that the Hamiltonian itself is an invariant tensor under $T_-$. Furthermore, observe that the Hamiltonian is fully invariant under
\begin{equation}
	S \equiv -\lambda H^U -\lambda^2E_-^U + E_+^U = \begin{pmatrix}0 & 0 & 0\\ 0& -\lambda & 1\\0 &-\lambda^2 & \lambda \end{pmatrix} ~,
\end{equation}
which is a linear combination of $U$-spin operators, and therefore must be QED charge preserving, too.

The invariance of $H$ under these operators is related to the generation of sum rules for the amplitudes. As in the main text, a sum rule itself has the form
\begin{equation}
	\mathcal{S}^{\alpha\beta\mu}(C_w)_{\alpha\beta\mu} = 0~.
\end{equation}
The index $\mu$ here is a tensor index. To see this, note that that the meson tensor $(D_\mu)^i = \partial (D_3)^i/ \partial D_\mu = \delta_\mu^i$. Then under an operator $T$,
\begin{equation}
	T |D_\mu\rangle = |D_\alpha\rangle\langle D_\alpha|T |D_\mu\rangle =  (D^\alpha)_jT^j_i (D_\mu)^i|D_\alpha\rangle = T^\alpha_\mu|D_\alpha\rangle~.
\end{equation} 
Similarly for the out-state mesons we just have $\langle M_\alpha | = (M_\alpha)^\rho_\sigma \langle M^\rho_\sigma |$, where $\langle M^\rho_\sigma |$ are a normalized basis of the out-states. The completeness relation $\sum_\alpha\mbox{Tr}\{A M^T_\alpha\} M_\alpha \equiv A - \bm{1}\mbox{Tr}\{A\}$ and the tracelessness of $T$ and $M_\alpha$ then implies
\begin{equation}
	T \langle M_\alpha | =  \mbox{Tr}\big\{[T,M_\alpha] M^T_\beta \big\}  \langle M_\beta | \equiv [T_8]^\beta_\alpha \langle M_\beta |~,
\end{equation}
so that we may treat $\alpha$ and $\beta$ as indices in a basis transforming under the adjoint representation.

The key observation in generating sum rules abstractly is that provided $TH = 0$, then it follows that $TC_w = 0$ by eq. (\ref{eqn:HXC}). Hence
\begin{equation}
	\label{eqn:MSR}
	T^{\rho\sigma\gamma}_{\alpha\beta\mu}[C_w]_{\rho\sigma\gamma} \equiv [T_8]^\gamma_\alpha [C_w]_{\gamma\beta\mu} + [T_8]^\gamma_\beta [C_w]_{\alpha\gamma\mu} + T^\gamma_\mu [C_w]_{\alpha\beta\gamma} = 0~.
\end{equation}
This master formula permits us to compute sum rules without computing the Wigner-Eckhart invariants: one need only select appropriate $\alpha$, $\beta$, $\mu$ to generate a sum rule. As an example of the operation of this master formula, let us now consider $T = S$ or $T_-$, the two operators under which $H$ is invariant. For $T=S$, we have in the $D \to P_8P_8$ case
\begin{align}
\langle M_\alpha | & = \begin{pmatrix} \langle \pi^0| & \langle \eta_8| & \langle \pi^+| & \langle K^+ | & \langle \pi^-| & \langle K^- | & \langle K^0| &\langle \bar{K}^0 | \end{pmatrix} \notag\\ 
[S_8]^\beta_\alpha & = \begin{pmatrix}
 0 & 0 & 0 & 0 & 0 & 0 & -\frac{\lambda ^2}{\sqrt{2}} & -\frac{1}{\sqrt{2}} \\
 0 & 0 & 0 & 0 & 0 & 0 & \sqrt{\frac{3}{2}} \lambda ^2 & \sqrt{\frac{3}{2}} \\
 0 & 0 & \lambda  & \lambda^2 & 0 & 0 & 0 & 0 \\
 0 & 0 & -1 & -\lambda  & 0 & 0 & 0 & 0 \\
 0 & 0 & 0 & 0 & -\lambda  & 1 & 0 & 0 \\
 0 & 0 & 0 & 0 & -\lambda^2 & \lambda  & 0 & 0 \\
 \frac{1}{\sqrt{2}} & -\sqrt{\frac{3}{2}} & 0 & 0 & 0 & 0 & -2 \lambda  & 0 \\
 \frac{\lambda ^2}{\sqrt{2}} & -\sqrt{\frac{3}{2}} \lambda^2 & 0 & 0 & 0 & 0 & 0 & 2 \lambda 
\end{pmatrix}~.
\end{align}
The operator $S$ does not change the electric charge of the states, so that provided each choice of $\alpha,\beta,\mu$ corresponds to a QED preserving amplitude $D_\mu \to [P_8]_\alpha [P_8]_\beta$, then eq. (\ref{eqn:MSR}) generates a sum-rule. For example, the linear combination of the choices $\{\alpha,\beta,\mu\} = \{K^+,\pi^-,D^0\}$ and $\{\alpha,\beta,\mu\} = \{K^-,\pi^+,D^0\}$ generates the U-spin sum rule
\begin{align}
	0 & = S^{\rho\sigma\gamma}_{K^+\pi^-D^0}[C_w]_{\rho\sigma\gamma}/2\lambda^3 +  S^{\rho\sigma\gamma}_{K^-\pi^+D^0}[C_w]_{\rho\sigma\gamma}/2\lambda \notag\\
	& = -\frac{[C_w]_{K^-K^+D^0}}{\lambda} + \frac{[C_w]_{\pi^-\pi^+D^0}}{\lambda} + [C_w]_{K^-\pi^+D^0} - \frac{[C_w]_{\pi^-K^+D^0}}{\lambda^2}~.
\end{align}
Similarly, for the operator $T_- \equiv E^I_- + \lambda E^D_-$ one has
\begin{equation}
	[T_{-8}]^\beta_\alpha = \begin{pmatrix}
 0 & 0 & -\sqrt{2} & -\frac{\lambda }{\sqrt{2}} & 0 & 0 & 0 & 0 \\
 0 & 0 & 0 & -\sqrt{\frac{3}{2}} \lambda  & 0 & 0 & 0 & 0 \\
 0 & 0 & 0 & 0 & 0 & 0 & 0 & 0 \\
 0 & 0 & 0 & 0 & 0 & 0 & 0 & 0 \\
 \sqrt{2} & 0 & 0 & 0 & 0 & 0 & -\lambda  & 0 \\
 \frac{\lambda }{\sqrt{2}} & \sqrt{\frac{3}{2}} \lambda  & 0 & 0 & 0 & 0 & 0 & -1 \\
 0 & 0 & 0 & 1 & 0 & 0 & 0 & 0 \\
 0 & 0 & \lambda  & 0 & 0 & 0 & 0 & 0
\end{pmatrix}~.
\end{equation}
The operator $T_-$ is a $\Delta Q = -1$ operator, so that eq. (\ref{eqn:MSR}) produces a sum rule of QED preserving amplitudes for each choice of $\alpha, \beta, \mu$ corresponding to a $\Delta Q = +1$ amplitude.

In order to produce the sum rules for our effective Hamiltonian (\ref{eqn:CKMH}), it is important to observe that not all QED preserving amplitudes can be produced by $H$. In particular, $H$ can raise or lower U-spin by at most one unit, so that the $\Delta U = \pm 2$ amplitudes
\begin{equation}
	A_{D^0 \to K^0K^0}~,\qquad A_{D^0 \to \bar{K}^0 \bar{K}^0}~,\qquad A_{D^+ \to K^0K^+}~,\qquad A_{D_s^+ \to \bar{K}^0\pi^+}~,
\end{equation} 
must be zero. Enforcing the zero value for these amplitudes, one obtains 21 linearly independent $P_8P_8$ sum-rules from the sum rule operators $S$ and $T_-$, in the SU(3) flavor basis

\setcounter{incre}{0}

\benum
\item
\begin{align*} -3 A_{D^0\to 2\eta_8} &+A_{D^0\to 2\pi_0} -4 \sqrt{6} \lambda  A_{D^0\to \eta_8  \bar{K}^0} +2 A_{D^0\to K^0 \bar{K}^0} =0     \end{align*}
 \item
\begin{align*}  \dfrac{3 A_{D^0\to 2\eta_8}}{\sqrt{2}} &+3 \sqrt{3} \lambda  A_{D^0\to \eta_8  \bar{K}^0} -\dfrac{3 A_{D^0\to K^0
   \bar{K}^0}}{\sqrt{2}}  -\dfrac{A_{D^+\to \bar{K}^0 K^+}}{\sqrt{2}} +A_{D^+\to \pi _0 \pi ^+} \\[10pt]
   & -\dfrac{\lambda  A_{D^+_s\to \bar{K}^0 K^+}}{\sqrt{2}} = 0         \end{align*}
 \item
\begin{align*} A_{D^+_s\to \pi _0 \pi ^+} & =0        \end{align*}
 \item
\begin{align*} 3 \sqrt{3} \lambda ^2 A_{D^0\to \eta_8  \bar{K}^0}  &+\dfrac{3 A_{D^0\to 2\eta_8} \lambda }{\sqrt{2}} -\dfrac{3 \lambda A_{D^0\to K^0 \bar{K}^0}  }{\sqrt{2}} -\dfrac{\lambda A_{D^+\to \bar{K}^0 K^+}  }{\sqrt{2}} +A_{D^+\to \pi _0 K^+} =0  \end{align*}
 \item
\begin{align*} -\dfrac{3 A_{D^0\to 2\eta_8}}{\sqrt{2}} & -3 \sqrt{3} \lambda  A_{D^0\to \eta_8  \bar{K}^0} +\dfrac{3
   A_{D^0\to K^0 \bar{K}^0}}{\sqrt{2}} +A_{D^+_s\to \pi _0 K^+} -\dfrac{\lambda  A_{D^+_s\to \bar{K}^0
   K^+}}{\sqrt{2}} =0  \end{align*}
 \item
\begin{align*} -\sqrt{3} A_{D^0\to 2\eta_8} +A_{D^0\to \eta_8  \pi _0} & -2 \sqrt{2} \lambda  A_{D^0\to \eta_8  \bar{K}^0} +\sqrt{3}
   A_{D^0\to K^0 \bar{K}^0}  = 0    \end{align*}
 \item
\begin{align*} \sqrt{3} \lambda ^2A_{D^0\to \eta_8  \bar{K}^0}  & +A_{D^0\to K^0 \pi _0}  = 0       \end{align*}
 \item
\begin{align*}  A_{D^0\to \pi _0 \bar{K}^0} & -\sqrt{3} A_{D^0\to \eta_8  \bar{K}^0}  = 0       \end{align*}
 \item
\begin{align*} -6 A_{D^0\to 2\eta_8} &-7 \sqrt{6} \lambda  A_{D^0\to \eta_8  \bar{K}^0} +5 A_{D^0\to K^0 \bar{K}^0} +A_{D^0\to \pi ^- \pi ^+} +A_{D^+\to \bar{K}^0 K^+} \\[10pt]
& +\lambda  A_{D^+_s\to \bar{K}^0 K^+} =0        \end{align*}
 \item
\begin{align*} -3  \sqrt{\dfrac{3}{2}} A_{D^0\to 2\eta_8} &-9 \lambda  A_{D^0\to \eta_8  \bar{K}^0} +3 \sqrt{\dfrac{3}{2}} A_{D^0\to K^0
   \bar{K}^0}     +\dfrac{A_{D^+\to \bar{K}^0 K^+}}{\sqrt{6}} +A_{D^+\to \eta_8  \pi ^+} \\[10pt]
 &   +\sqrt{\dfrac{3}{2}} \lambda A_{D^+_s\to \bar{K}^0 K^+} =0      \end{align*}
 \item
\begin{align*} -\dfrac{\sqrt{6} A_{D^0\to 2\eta_8}}{\lambda } &-6 A_{D^0\to \eta_8  \bar{K}^0} +\dfrac{\sqrt{6} A_{D^0\to K^0 \bar{K}^0}}{\lambda } +\sqrt{\dfrac{2}{3}}\dfrac{ A_{D^+\to \bar{K}^0 K^+}}{\lambda } +A_{D^+_s\to \eta_8  \pi ^+} =0   \end{align*}
 \item
\begin{align*} \lambda ^2A_{D^+_s\to \bar{K}^0 K^+}  & +A_{D^+\to K^0 \pi ^+} =0        \end{align*}
 \item
\begin{align*}-6 A_{D^0\to 2\eta_8}& -6 \sqrt{6} \lambda  A_{D^0\to \eta_8  \bar{K}^0} +6 A_{D^0\to K^0 \bar{K}^0} +A_{D^+\to \bar{K}^0
   K^+} +A_{D^+_s\to K^0 \pi ^+} =0   \end{align*}
 \item
\begin{align*} \dfrac{3 A_{D^0\to 2\eta_8}}{\lambda }& +4 \sqrt{6} A_{D^0\to \eta_8  \bar{K}^0} -\dfrac{3 A_{D^0\to K^0 \bar{K}^0}}{\lambda } +A_{D^0\to K^- \pi ^+} -\dfrac{A_{D^+\to \bar{K}^0 K^+}}{\lambda } \\[10pt]
 &-A_{D^+_s\to \bar{K}^0 K^+} =0          \end{align*}
 \item
\begin{align*} \dfrac{3 A_{D^0\to 2\eta_8}}{\lambda } &+3 \sqrt{6} A_{D^0\to \eta_8  \bar{K}^0} -\dfrac{3 A_{D^0\to K^0 \bar{K}^0}}{\lambda } -\dfrac{A_{D^+\to \bar{K}^0 K^+}}{\lambda } +A_{D^+\to \bar{K}^0 \pi ^+} \\[10pt] 
&-A_{D^+_s\to \bar{K}^0 K^+} =0         \end{align*}
 \item
\begin{align*} -4 \sqrt{6} \lambda ^2A_{D^0\to \eta_8  \bar{K}^0} & +\lambda ^2A_{D^+_s\to \bar{K}^0 K^+}  -3\lambda A_{D^0\to 2\eta_8}     +3\lambda A_{D^0\to K^0 \bar{K}^0}   +\lambda A_{D^+\to \bar{K}^0 K^+}   \\[10pt]
  &   +A_{D^0\to \pi ^- K^+} =0         \end{align*}
 \item
\begin{align*}-3 \lambda ^2A_{D^0\to \eta_8 \bar{K}^0} & -\sqrt{\dfrac{3}{2}}\lambda A_{D^0\to 2\eta_8}   +\sqrt{\dfrac{3}{2}} \lambda A_{D^0\to K^0 \bar{K}^0}   +\dfrac{\lambda A_{D^+\to \bar{K}^0 K^+}  }{\sqrt{6}} +A_{D^+\to \eta_8  K^+} =0   \end{align*}
 \item
\begin{align*}-\sqrt{\dfrac{3}{2}} A_{D^0\to 2\eta_8} &-3 \lambda  A_{D^0\to \eta_8  \bar{K}^0} +\sqrt{\dfrac{3}{2}} A_{D^0\to K^0 \bar{K}^0}  +\sqrt{\dfrac{2}{3}} A_{D^+\to \bar{K}^0 K^+} +A_{D^+_s\to \eta_8  K^+} \\[10pt]
 &   +\sqrt{\dfrac{3}{2}} \lambda  A_{D^+_s\to \bar{K}^0 K^+} =0          \end{align*}
 \item
\begin{align*}-3 \sqrt{6}  \lambda ^2A_{D^0\to \eta_8  \bar{K}^0} &+\lambda^2A_{D^+_s\to \bar{K}^0 K^+}  -3 \lambda A_{D^0\to 2\eta_8}  +3\lambda A_{D^0\to K^0 \bar{K}^0}   +\lambda A_{D^+\to \bar{K}^0 K^+}   \\[10pt] 
 & +A_{D^+_s\to K^0 K^+} =0         \end{align*}
 \item
\begin{align*} \sqrt{6} \lambda A_{D^0\to \eta_8  \bar{K}^0}& -A_{D^0\to K^0 \bar{K}^0} +A_{D^0\to K^- K^+} -A_{D^+\to \bar{K}^0 K^+} -\lambda  A_{D^+_s\to \bar{K}^0 K^+} =0   \end{align*}
 \item
\begin{align*} \lambda ^2A_{D^0\to \eta_8  \bar{K}^0} & +A_{D^0\to \eta_8  K^0}=0~.    \end{align*}
\eenum

In the $D \to P_1P_8$ case, we have final states $\langle \eta_1 M_\beta|$, where $\langle \eta_1|$ is an SU(3) singlet, so that the master formula (\ref{eqn:MSR}) becomes
\begin{equation}
	\label{eqn:MSR1}
	T^{\sigma\gamma}_{\beta\mu}[C_w]_{\eta_1\sigma\gamma} \equiv  [T_8]^\gamma_\beta [C_w]_{\eta_1\gamma\mu} + T^\gamma_\mu [C_w]_{\eta_1\beta\gamma} = 0~.
\end{equation}
Applying eq. (\ref{eqn:MSR1}) to all possible $\beta$,$\mu$ that correspond to QED-preserving $|\Delta U| < 2$ amplitudes, one further finds 5 linearly independent  $P_1P_8$ sum rules. In the SU(3) flavor basis these are

\benum
\item
\begin{align*}
 -\frac{\sqrt{3} A_{D^0\to \eta  \eta_1}}{\lambda } & +\frac{A_{D^0\to \pi _0 \eta_1}}{\lambda }-2 \sqrt{2} A_{D^0\to \bar{K}^0 \eta_1} = 0
\end{align*}
\item
\begin{align*}
 -\frac{2 \sqrt{6}
   A_{D^0\to \eta  \eta_1}}{\lambda }-6 A_{D^0\to \bar{K}^0 \eta_1}& +\frac{A_{D^+\to \pi ^+ \eta_1}}{\lambda }+\frac{A_{D^+_s\to K^+ \eta 
   '}}{\lambda } =0
\end{align*}
\item
\begin{align*}
 \frac{\sqrt{6} A_{D^0\to \eta  \eta_1}}{\lambda }& +3 A_{D^0\to \bar{K}^0 \eta_1}-\frac{A_{D^+_s\to K^+ \eta_1}}{\lambda
   }+A_{D^+_s\to \pi ^+ \eta_1}  =0
\end{align*}
\item
\begin{align*}
 -\frac{\sqrt{6} A_{D^0\to \eta  \eta_1}}{\lambda }& -3 A_{D^0\to \bar{K}^0 \eta_1}+\frac{A_{D^+\to K^+
   \eta_1}}{\lambda ^2}+\frac{A_{D^+_s\to K^+ \eta_1}}{\lambda } =0
\end{align*}
\item
\begin{align*}
 \frac{A_{D^0\to K^0 \eta_1}}{\lambda ^2}& +A_{D^0\to \bar{K}^0 \eta_1}=0~.
\end{align*}
\eenum

These 26 sum rules coincide precisely with those found by direct computation, and can be verified with by reference to the tables of Appendix \ref{app:PP}. Similar results can be obtained for the PV case. This method of computing sum rules applies only in the case that the invariants can be written in terms of the Hamiltonian (cf. eq. (\ref{eqn:HXC})) and there exist operators under which the Hamiltonian is invariant. Once we introduce the s-mass spurion then observe $T_- m_s$, $S m_s \not= 0$, so the above analysis fails at first order in the spurion, unless one finds an operator under which both $H$ and $m_s$ are invariants. Unfortunately, there does not seem to be such an operator available, so we are left with the option of just computing the invariants directly. 

\clearpage 

\section{\texorpdfstring{$D \to PP$}{DPP} Invariants}
\label{app:PP}
\subsection{\texorpdfstring{$\mathcal{O}(1)$}{O1} Invariants}
There are in principle seven linearly independent $P_8P_8$ Wigner-Eckhart invariants at $\mathcal{O}(1)$,  four invariants for the $P_1P_8$ case and two for the $P_1P_1$ case. Due to the proportionality of the two $\bm{\bar{3}}$ irreps, these are reduced respectively to five, two and one linearly independent invariants. The $\mathcal{O}(1)$ invariants are shown in the tables below, for the $P_1P_1$, $P_1P_8$, and $P_8P_8$ amplitudes, labelled by $w = [_{xy}^{\rm pp}]^k_R$, where $x,y$=$1,8$ labels the P representations, $R$ is the $H$ irrep generating the invariant, and $k$ indicates the $k$th such invariant. For convenience, we write only the invariant subscripts, so that
\begin{equation}
	\label{eqn:AIL}
	C_{[_{xy}^{\rm pp}]^k_R} = [_{xy}^{\rm pp}]^k_R~.
\end{equation}

Now, the $\mathcal{O}(1)$ invariants are
\begin{gather}
	[_{88}^{\rm pp}]_{\bm{\bar{3}}}^1 = [M_P]^i_j[M_P]^j_i \big(3[\bm{\bar{3}'}]_k -[\bm{\bar{3}}]_k\big)  [D]^k/8~,\quad  [_{88}^{\rm pp}]_{\bm{\bar{3}}}^2 =  [M_P]^i_j[M_P]^j_k \big(3[\bm{\bar{3}'}]_i-  [\bm{\bar{3}}]_i\big) [D]^k/8~,\notag\\
	[_{88}^{\rm pp}]_{\bm{6}} = [M_P]^i_j[M_P]^k_i [\bm{6}]^{lj} [D]^m\varepsilon_{klm}~,\notag\\
	[_{88}^{\rm pp}]_{\bm{\bar{15}}}^1 = [M_P]^i_j[M_P]^j_k [\bm{\bar{15}}]^k_{il} [D]^l~,[_{88}^{\rm pp}]_{\bm{\bar{15}}}^2= [M_P]^i_j[M_P]^k_l [H_{15}]^j_{ik} [D]^l~,\notag\\
	[_{18}^{\rm pp}]_{\bm{\bar{3}}} = \eta_1[M_P]^j_i \big(3[\bm{\bar{3}'}]_j -[\bm{\bar{3}}]_j\big) [D]^i/8~,\notag\\
	[_{18}^{\rm pp}]_{\bm{6}} = \eta_1[M_P]^i_j [\bm{6}]^{lj} [D]^m\varepsilon_{ilm}~,\quad [_{18}^{\rm pp}]_{\bm{\bar{15}}} = \eta_1 [M_P]^i_j[H_{15}]^j_{ik} [D]^k~,\notag\\
	[_{11}^{\rm pp}]_{\bm{\bar{3}}} = \eta_1\eta_1 \big(3[\bm{\bar{3}'}]_i -[\bm{\bar{3}}]_i\big) [D]^i/8~.
\end{gather}
These invariants are shown explicitly in Tables \ref{tab:PP1118} and \ref{tab:PP88}. Of the 39 possible QED preserving amplitudes, only 34 are non-zero. The remaining five zero amplitudes are
\begin{equation}
	A_{D^0 \to K^0K^0}~,\qquad A_{D^0 \to \bar{K}^0 \bar{K}^0}~,\qquad A_{D^+ \to K^0K^+}~,\qquad A_{D_s^+ \to \bar{K}^0\pi^+}~,\qquad A_{D_s^+ \to \pi_0\pi^+}~,
\end{equation} 
and therefore are not shown in the tables. Of these, the first four are the $\Delta U = \pm 2$ amplitudes; the fifth amplitude is accidentally zero, as predicted in the sum rule (iii) of Appendix \ref{app:ASRG}.
 
\begin{table}[ht]
\begin{center}
\begin{tabular*}{0.75\linewidth}{| >{$}c<{$}  >{$}c<{$} |@{\extracolsep{\fill}} >{$}c<{$} >{$}c<{$} >{$}c<{$} >{$}c<{$} |}
\hline
\Delta U & \mbox{\small{Ampl.}} & [_{11}^{\rm pp}]_{\bm{\bar{3}}}  & [_{18}^{\rm pp}]_{\bm{\bar{3}}}   & [_{18}^{\rm pp}]_{\bm{6}}  &   [_{18}^{\rm pp}]_{\bm{\bar{15}}}\\
\hline
\hline
 0 & \sml{D^0\to 2\eta_1} & \frac{1}{4} \lambda^5 A^2 & 0 & 0 & 0 \\
 0 & \sml{D^0\to \eta_1 \eta_8} & 0 & \frac{\lambda^5 A^2}{8 \sqrt{6}} & -\frac{\lambda}{2} \sqrt{\frac{3}{2}}  & -\frac{\lambda}{2} \sqrt{\frac{3}{2}} 
   \\
 1 & \sml{D^0\to K^0 \eta_1} & 0 & 0 & -\frac{\lambda^2}{2} & -\frac{\lambda^2}{2} \\
 0 & \sml{D^0\to \pi_0 \eta_1} & 0 & \frac{\lambda^5 A^2}{8 \sqrt{2}} & \frac{\lambda }{2 \sqrt{2}} & \frac{\lambda }{2 \sqrt{2}} \\
 -1 & \sml{D^0\to \eta_1 \bar{K}^0} & 0 & 0 & \frac{1}{2} & \frac{1}{2} \\
 \hline
 1 & \sml{D^+\to \eta_1 K^+} & 0 & 0 & \frac{\lambda^2}{2} & -\frac{\lambda^2}{2} \\
 0 & \sml{D_s^+\to \eta_1 K^+} & 0 & \frac{\lambda^5 A^2}{8}  & -\frac{\lambda }{2} & \frac{\lambda }{2} \\
 0 & \sml{D^+\to \eta_1 \pi^+} & 0 & \frac{\lambda^5 A^2}{8} & \frac{\lambda }{2} & -\frac{\lambda }{2} \\
 -1 & \sml{D_s^+\to \eta_1 \pi^+} & 0 & 0 & -\frac{1}{2} & \frac{1}{2}\\
 \hline
\end{tabular*}
\end{center}
\caption{$D \to P_1P_1$ and $D \to P_1P_8$ $\mathcal{O}(1)$ invariants. $A^2$ is shorthand for $A^2(\rho - i\eta)$.}
\label{tab:PP1118}
\end{table}

\begin{table}[ht]
\begin{center}
\begin{tabular*}{0.9\linewidth}{| >{$}c<{$}  >{$}c<{$} | @{\extracolsep{\fill}}>{$}c<{$} >{$}c<{$} >{$}c<{$} >{$}c<{$} >{$}c<{$} |}
\hline
\Delta U & \mbox{\small{Ampl.}} & [_{88}^{\rm pp}]_{\bm{\bar{3}}}^1 & [_{88}^{\rm pp}]_{\bm{\bar{3}}}^2  & [_{88}^{\rm pp}]_{\bm{6}} & [_{88}^{\rm pp}]_{\bm{\bar{15}}}^1 &[_{88}^{\rm pp}]_{\bm{\bar{15}}}^2\\
\hline
\hline
0 & \sml{D^0\to \pi^- \pi^+} & 0 &  \frac{\lambda^5 A^2}{4} & -\frac{\lambda }{2} & -\lambda  & -\frac{\lambda }{2} \\
 -1 & \sml{D^0\to K^- \pi^+} & 0 & 0 & \frac{1}{2} & 1 & \frac{1}{2} \\
  1 & \sml{D^0\to \pi^- K^+} & 0 & 0 & -\frac{\lambda^2}{2} & -\lambda^2 & -\frac{\lambda^2}{2} \\
   0 & \sml{D^0\to K^- K^+} & 0 &  \frac{\lambda^5 A^2}{4} & \frac{\lambda }{2} & \lambda  & \frac{\lambda }{2} \\
\hline
    0 & \sml{D^0\to 2\eta_8} & - \frac{\lambda^5 A^2}{6} &  \frac{\lambda^5 A^2}{4} & \frac{\lambda }{2} & \lambda  & -\frac{\lambda }{2} \\
 1 & \sml{D^0\to K^0 \eta_8} & 0 & 0 & \frac{\lambda^2}{2 \sqrt{6}} & \frac{\lambda^2}{\sqrt{6}} & -\frac{\lambda^2}{2 \sqrt{6}} \\
 -1 & \sml{D^0\to \eta_8 \bar{K}^0} & 0 & 0 & -\frac{1}{2 \sqrt{6}} & -\frac{1}{\sqrt{6}} & \frac{1}{2 \sqrt{6}} \\
 0 & \sml{D^0\to K^0 \bar{K}^0} & - \frac{\lambda^5 A^2}{4} &  \frac{\lambda^5 A^2}{4} & 0 & 0 & 0\\
 \hline
  0 & \sml{D^0\to 2\pi_0} & 0 &  \frac{\lambda^5 A^2}{4} & -\frac{\lambda }{2} & -\lambda  & \frac{\lambda }{2} \\
 0 & \sml{D^0\to \pi_0 \eta_8} & \frac{\lambda^5 A^2}{4 \sqrt{3}} & 0 & \frac{\lambda }{2 \sqrt{3}} & \frac{\lambda }{\sqrt{3}} & -\frac{\lambda }{2
   \sqrt{3}} \\
 1 & \sml{D^0\to K^0 \pi_0} & 0 & 0 & \frac{\lambda^2}{2 \sqrt{2}} & \frac{\lambda^2}{\sqrt{2}} & -\frac{\lambda^2}{2 \sqrt{2}} \\
-1 & \sml{D^0\to \pi_0 \bar{K}^0} & 0 & 0 & -\frac{1}{2 \sqrt{2}} & -\frac{1}{\sqrt{2}} & \frac{1}{2 \sqrt{2}} \\
\hline
 0 & \sml{D^+\to \pi_0 \pi^+} & 0 & 0 & 0 & 0 & \frac{\lambda }{\sqrt{2}} \\
 1 & \sml{D^+\to \pi_0 K^+} & 0 & 0 & \frac{\lambda^2}{2 \sqrt{2}} & -\frac{\lambda^2}{\sqrt{2}} & \frac{\lambda^2}{2 \sqrt{2}} \\
 0 & \sml{D_s^+\to \pi_0 K^+} & \frac{\lambda^5 A^2}{4 \sqrt{2}} & 0 & -\frac{\lambda }{2 \sqrt{2}} & \frac{\lambda }{\sqrt{2}} &
   \frac{\lambda }{2 \sqrt{2}} \\
  0 & \sml{D^+\to \eta_8 \pi^+} & \frac{\lambda^5 A^2}{2 \sqrt{6}} & 0 & \frac{\lambda }{\sqrt{6}} & -\sqrt{\frac{2}{3}} \lambda  & -\sqrt{\frac{2}{3}}
   \lambda  \\
 -1 & \sml{D_s^+\to \eta_8 \pi^+} & 0 & 0 & -\frac{1}{\sqrt{6}} & \sqrt{\frac{2}{3}} & -\frac{1}{\sqrt{6}} \\
 1 & \sml{D^+\to K^0 \pi^+} & 0 & 0 & \frac{\lambda^2}{2} & -\lambda^2 & -\frac{\lambda^2}{2} \\
 0 & \sml{D_s^+\to K^0 \pi^+} &  \frac{\lambda^5 A^2}{4} & 0 & -\frac{\lambda }{2} & \lambda  & -\frac{\lambda }{2} \\
 -1 & \sml{D^+\to \bar{K}^0 \pi^+} & 0 & 0 & 0 & 0 & 1 \\
 1 & \sml{D^+\to \eta_8 K^+} & 0 & 0 & -\frac{\lambda^2}{2 \sqrt{6}} & \frac{\lambda^2}{\sqrt{6}} & -\frac{\lambda^2}{2 \sqrt{6}} \\
 0 & \sml{D_s^+\to \eta_8 K^+} & -\frac{\lambda^5 A^2}{4 \sqrt{6}} & 0 & \frac{\lambda }{2 \sqrt{6}} & -\frac{\lambda }{\sqrt{6}} & -\frac{5
   \lambda }{2 \sqrt{6}} \\
 1 & \sml{D_s^+\to K^0 K^+} & 0 & 0 & 0 & 0 & -\lambda^2 \\
 0 & \sml{D^+\to \bar{K}^0 K^+} &  \frac{\lambda^5 A^2}{4} & 0 & \frac{\lambda }{2} & -\lambda  & \frac{\lambda }{2} \\
 -1 & \sml{D_s^+\to \bar{K}^0 K^+} & 0 & 0 & -\frac{1}{2} & 1 & \frac{1}{2} \\
 \hline
\end{tabular*}
\end{center}
\caption{$D \to P_8P_8$ $\mathcal{O}(1)$ invariants. $A^2$ is shorthand for $A^2(\rho - i\eta)$.}
\label{tab:PP88}
\end{table}

\clearpage

\subsection{Spurionic \texorpdfstring{$\mathcal{O}(\varepsilon)$}{OE} Invariants}
The invariants produced by $Hm_s$ are shown in the following Tables \ref{tab:PP1118S}, \ref{tab:PP88S3}, \ref{tab:PP88S6} and \ref{tab:PP88S15}. In this case, the number of invariants increases dramatically, so that for the sake of brevity we do not list the explicit contractions corresponding to each invariant. Each invariant is labelled as in the previous section, but with an extra $s$ subscript. 

\begin{table}[ht]
\begin{center}
\begin{tabular*}{0.7\linewidth}{|>{$}c<{$}  >{$}c<{$} |@{\extracolsep{\fill}}>{$}c<{$} >{$}c<{$} >{$}c<{$}|}
\hline
\Delta U & \bm{\times~ \varepsilon} & [_{11}^{\rm pp}]_{\bm{\bar{3}}s}  &  [_{11}^{\rm pp}]_{\bm{6}s} &  [_{11}^{\rm pp}]_{\bm{\bar{15}}s}\\
\hline
\hline
0  &  \sml{D^0\to 2\eta_1}  &  \frac{\Delta}{4}   &  -3  \lambda   &  -3  \lambda \\
\hline
\end{tabular*}
\vspace{1cm} 

\begin{tabular*}{0.7\linewidth}{|>{$}c<{$}  >{$}c<{$} |@{\extracolsep{\fill}}>{$}c<{$} >{$}c<{$} >{$}c<{$}|}
\hline
\Delta U & \bm{\times~ \varepsilon\Delta} &  [_{18}^{\rm pp}]_{\bm{\bar{3}}s}^{1} & [_{18}^{\rm pp}]_{\bm{\bar{3}}s}^{2} &  [_{18}^{\rm pp}]_{\bm{\bar{3}}s}^{3} \\
\hline
\hline
 0 & \sml{D^0\to \pi_0 \eta_1} & \frac{1}{4 \sqrt{2}} & \frac{1}{8 \sqrt{2}} & \frac{1}{8 \sqrt{2}} \\
 0 & \sml{D^0\to \eta_1 \eta_8} & -\frac{1}{2 \sqrt{6}} & \frac{1}{8 \sqrt{6}} & \frac{1}{8 \sqrt{6}}\\
 \hline
 0 & \sml{D^+\to \eta_1 \pi^+} & \frac{1}{4} & \frac{1}{8} & \frac{1}{8} \\
 0 & \sml{D_s^+\to \eta_1 K^+} & -\frac{1}{8} & \frac{1}{8} & -\frac{1}{4} \\
\hline
\end{tabular*}
\vspace{1cm} 

\begin{tabular*}{\linewidth}{|>{$}c<{$}  >{$}c<{$} |@{\extracolsep{\fill}}>{$}c<{$} >{$}c<{$} >{$}c<{$}>{$}c<{$} >{$}c<{$} >{$}c<{$} >{$}c<{$}|}
\hline
\Delta U & \bm{\times~ \varepsilon} &  [_{18}^{\rm pp}]_{\bm{6}s}^{1} & [_{18}^{\rm pp}]_{\bm{6}s}^{2} &  [_{18}^{\rm pp}]_{\bm{6}s}^{3} &  [_{18}^{\rm pp}]_{\bm{\bar{15}}s}^{1} &  [_{18}^{\rm pp}]_{\bm{\bar{15}}s}^{2} &  [_{18}^{\rm pp}]_{\bm{\bar{15}}s}^{3} &  [_{18}^{\rm pp}]_{\bm{\bar{15}}s}^{4}\\
\hline
\hline
 0 & \sml{D^0\to \pi_0 \eta_1} & \frac{\lambda }{2 \sqrt{2}} & -\frac{\lambda }{\sqrt{2}} & -\frac{\lambda }{2 \sqrt{2}} & -\frac{3 \lambda }{2 \sqrt{2}} &
   -\frac{\lambda }{\sqrt{2}} & -\frac{3 \lambda }{2 \sqrt{2}} & \frac{\lambda }{2 \sqrt{2}} \\
    0 & \sml{D^0\to \eta_1 \eta_8} & \frac{\lambda}{2} \sqrt{\frac{3}{2}}  & 0 & \frac{\lambda}{2} \sqrt{\frac{3}{2}}  & \frac{ \lambda}{2} \sqrt{\frac{3}{2}}
    & -\sqrt{\frac{3}{2}} \lambda  & -\frac{\lambda}{2} \sqrt{\frac{3}{2}}  & \frac{\lambda}{2} \sqrt{\frac{3}{2}}  \\
 1 & \sml{D^0\to K^0 \eta_1} & \lambda^2 & \lambda^2 & \frac{\lambda^2}{2} & 0 & -\frac{\lambda^2}{2} & 0 & -\frac{\lambda^2}{2} \\
 -1 & \sml{D^0\to \eta_1 \bar{K}^0} & \frac{1}{2} & \frac{1}{2} & -\frac{1}{2} & -\frac{3}{2} & \frac{1}{2} & 0 & -1\\
\hline
 0 & \sml{D^+\to \eta_1 \pi^+} & \frac{\lambda }{2} & -\lambda  & -\frac{\lambda }{2} & -\frac{3 \lambda }{2} & \lambda  & -\frac{3 \lambda }{2} &
   -\frac{\lambda }{2} \\
 -1 & \sml{D_s^+\to \eta_1 \pi^+} & -\frac{1}{2} & -\frac{1}{2} & -1 & \frac{3}{2} & -1 & 0 & \frac{1}{2} \\
 1 & \sml{D^+\to \eta_1 K^+} & -\lambda^2 & -\lambda^2 & -\frac{\lambda^2}{2} & 0 & -\frac{\lambda^2}{2} & 0 & -\frac{\lambda^2}{2} \\
 0 & \sml{D_s^+\to \eta_1 K^+} & \lambda  & -\frac{\lambda }{2} & -\lambda  & 0 & \frac{\lambda }{2} & -\frac{3 \lambda }{2} & \frac{\lambda }{2}
   \\
\hline
\end{tabular*}
\end{center}
\caption{$D \to P_1P_1$ and $D \to P_1P_8$ invariants at first order in spurion. $\Delta = \lambda^5A^2(\rho - i\eta)$.}
\label{tab:PP1118S}
\end{table}

\begin{table}[ht]
\begin{center}
\begin{tabular*}{0.8\linewidth}{|>{$}c<{$}  >{$}c<{$} |@{\extracolsep{\fill}}>{$}c<{$} >{$}c<{$} >{$}c<{$}>{$}c<{$} >{$}c<{$} |}
\hline
\Delta U &\bm{\times~ \varepsilon  \Delta}  & [_{88}^{\rm pp}]_{\bm{\bar{3}}s}^{1} &  [_{88}^{\rm pp}]_{\bm{\bar{3}}s}^{2} &  [_{88}^{\rm pp}]_{\bm{\bar{3}}s}^{3} &  [_{88}^{\rm pp}]_{\bm{\bar{3}}s}^{4} &  [_{88}^{\rm pp}]_{\bm{\bar{3}}s}^{5} \\
\hline
\hline
 0 & \sml{D^0\to \pi^- \pi^+} & 0 & 0 & \frac{1}{2} & \frac{1}{4} & 0 \\
 0 & \sml{D^0\to K^- K^+} & 0 & 0 & -\frac{1}{4} & \frac{1}{4} & 0 \\
\hline
 0 & \sml{D^0\to 2\pi_0} & 0 & 0 & \frac{1}{2} & \frac{1}{4} & 0 \\
 0 & \sml{D^0\to \pi_0 \eta_8} & \frac{1}{4 \sqrt{3}} & \frac{1}{4 \sqrt{3}} & 0 & 0 & \frac{\sqrt{3}}{8} \\
  0 & \sml{D^0\to 2\eta_8} & -\frac{1}{6} & -\frac{1}{6} & -\frac{1}{2} & \frac{1}{4} & \frac{1}{4} \\
 0 & \sml{D^0\to K^0 \bar{K}^0} & -\frac{1}{4} & -\frac{1}{4} & -\frac{1}{4} & \frac{1}{4} & 0\\
\hline
0 & \sml{D_s^+\to \pi_0 K^+} & \frac{1}{4 \sqrt{2}} & -\frac{1}{2 \sqrt{2}} & 0 & 0 & 0 \\0 & \sml{D^+\to \eta_8 \pi^+} & \frac{1}{2 \sqrt{6}} & \frac{1}{2 \sqrt{6}} & 0 & 0 & \frac{\sqrt{\frac{3}{2}}}{4} \\
 0 & \sml{D_s^+\to K^0 \pi^+} & \frac{1}{4} & -\frac{1}{2} & 0 & 0 & 0 \\
 0 & \sml{D_s^+\to \eta_8 K^+} & -\frac{1}{4 \sqrt{6}} & \frac{1}{2 \sqrt{6}} & 0 & 0 & \frac{\sqrt{\frac{3}{2}}}{4} \\
 0 & \sml{D^+\to \bar{K}^0 K^+} & \frac{1}{4} & \frac{1}{4} & 0 & 0 & 0 \\
\hline
\end{tabular*}
\end{center}
\caption{$D \to P_8P_8$ invariants at first order in spurion, generated by $\bm{\bar{3}}$. $\Delta = \lambda^5A^2(\rho - i\eta)$.}
\label{tab:PP88S3}
\end{table}

\begin{table}[t]
\begin{tabular*}{\linewidth}{|>{$}c<{$}  >{$}c<{$} |@{\extracolsep{\fill}}>{$}c<{$} >{$}c<{$} >{$}c<{$}>{$}c<{$} >{$}c<{$} >{$}c<{$} |}
\hline
\Delta U & \bm{\times~\varepsilon} &   [_{88}^{\rm pp}]_{\bm{6}s}^{1} & [_{88}^{\rm pp}]_{\bm{6}s}^{2} & [_{88}^{\rm pp}]_{\bm{6}s}^{3} & [_{88}^{\rm pp}]_{\bm{6}s}^{4} & [_{88}^{\rm pp}]_{\bm{6}s}^{5} & [_{88}^{\rm pp}]_{\bm{6}s}^{6} \\
\hline
\hline 
 0 & \sml{D^0\to \pi^- \pi^+} & 2 \lambda  & 0 & -\frac{\lambda }{2} & -\lambda  & 0 & 0 \\ 
  -1 & \sml{D^0\to K^- \pi^+} & -2 & 0 & \frac{1}{2} & -\frac{1}{2} & 0 & 0 \\
   1 & \sml{D^0\to \pi^- K^+} & -\lambda^2 & 0 & \lambda^2 & -\lambda^2 & 0 & 0 \\
    0 & \sml{D^0\to K^- K^+} & \lambda  & 0 & -\lambda  & -\frac{\lambda }{2} & 0 & 0 \\
\hline
 0 & \sml{D^0\to 2\pi_0} & 2 \lambda  & 0 & -\frac{\lambda }{2} & -\lambda  & 0 & 0 \\
 0 & \sml{D^0\to \pi_0 \eta_8} & \frac{\lambda }{\sqrt{3}} & -\frac{\sqrt{3} \lambda }{2} & -\frac{\lambda }{\sqrt{3}} & -\frac{\lambda }{2 \sqrt{3}} &
   \frac{\sqrt{3} \lambda }{2} & \frac{\sqrt{3} \lambda }{2} \\
 1 & \sml{D^0\to K^0 \pi_0} & \frac{\lambda^2}{\sqrt{2}} & 0 & -\frac{\lambda^2}{\sqrt{2}} & \frac{\lambda^2}{\sqrt{2}} & 0 & 0 \\
 -1 & \sml{D^0\to \pi_0 \bar{K}^0} & \sqrt{2} & 0 & -\frac{1}{2 \sqrt{2}} & \frac{1}{2 \sqrt{2}} & 0 & -\frac{3}{2 \sqrt{2}} \\
\hline
     0 & \sml{D^0\to 2\eta_8} & 0 & \lambda  & -\frac{\lambda }{2} & 0 & -3 \lambda  & -\lambda  \\
 1 & \sml{D^0\to K^0 \eta_8} & \frac{\lambda^2}{\sqrt{6}} & 0 & -\frac{\lambda^2}{\sqrt{6}} & \frac{\lambda^2}{\sqrt{6}} & -\sqrt{\frac{3}{2}} \lambda^2 &
   -\sqrt{\frac{3}{2}} \lambda^2 \\
 -1 & \sml{D^0\to \eta_8 \bar{K}^0} & \sqrt{\frac{2}{3}} & 0 & -\frac{1}{2 \sqrt{6}} & \frac{1}{2 \sqrt{6}} & \sqrt{\frac{3}{2}} & \sqrt{\frac{3}{8}}
   \\
 0 & \sml{D^0\to K^0 \bar{K}^0} & 0 & \frac{3 \lambda }{2} & 0 & 0 & 0 & \frac{3 \lambda }{2}\\
\hline
 1 & \sml{D^+\to \pi_0 K^+} & \frac{\lambda^2}{\sqrt{2}} & 0 & -\frac{\lambda^2}{\sqrt{2}} & \frac{\lambda^2}{\sqrt{2}} & 0 & 0 \\
 0 & \sml{D_s^+\to \pi_0 K^+} & -\frac{\lambda }{\sqrt{2}} & -\frac{3 \lambda }{2 \sqrt{2}} & -\frac{\lambda }{2 \sqrt{2}} & -\frac{\lambda
   }{\sqrt{2}} & 0 & 0 \\
 0 & \sml{D^+\to \eta_8 \pi^+} & \sqrt{\frac{2}{3}} \lambda  & -\sqrt{\frac{3}{2}} \lambda  & -\sqrt{\frac{2}{3}} \lambda  & -\frac{\lambda }{\sqrt{6}} &
   \sqrt{\frac{3}{2}} \lambda  & \sqrt{\frac{3}{2}} \lambda  \\
 -1 & \sml{D_s^+\to \eta_8 \pi^+} & -\sqrt{\frac{2}{3}} & 0 & -\frac{1}{\sqrt{6}} & \frac{1}{\sqrt{6}} & -\sqrt{\frac{3}{2}} & 0 \\
 1 & \sml{D^+\to K^0 \pi^+} & \lambda^2 & 0 & -\lambda^2 & \lambda^2 & 0 & 0 \\
 0 & \sml{D_s^+\to K^0 \pi^+} & -\lambda  & -\frac{3 \lambda }{2} & -\frac{\lambda }{2} & -\lambda  & 0 & 0 \\
 -1 & \sml{D^+\to \bar{K}^0 \pi^+} & 0 & 0 & 0 & 0 & 0 & -\frac{3}{2} \\
 1 & \sml{D^+\to \eta_8 K^+} & -\frac{\lambda^2}{\sqrt{6}} & 0 & \frac{\lambda^2}{\sqrt{6}} & -\frac{\lambda^2}{\sqrt{6}} & \sqrt{\frac{3}{2}} \lambda^2 &
   \sqrt{\frac{3}{2}} \lambda^2 \\
 0 & \sml{D_s^+\to \eta_8 K^+} & \frac{\lambda }{\sqrt{6}} & \frac{\lambda}{2} \sqrt{\frac{3}{2}}  & \frac{\lambda }{2 \sqrt{6}} &
   \frac{\lambda }{\sqrt{6}} & -\sqrt{\frac{3}{2}} \lambda  & 0 \\
 0 & \sml{D^+\to \bar{K}^0 K^+} & \lambda  & -\frac{3 \lambda }{2} & -\lambda  & -\frac{\lambda }{2} & 0 & -\frac{3 \lambda }{2} \\
 -1 & \sml{D_s^+\to \bar{K}^0 K^+} & -1 & 0 & -\frac{1}{2} & \frac{1}{2} & 0 & 0 \\
\hline
\end{tabular*}
\caption{$D \to P_8P_8$ invariants at first order in spurion, generated by $\bm{6}$.}
\label{tab:PP88S6}
\end{table}

\begin{table}[t]
\resizebox{\linewidth}{!}
{
\begin{tabular*}{1.1\linewidth}{|>{$}c<{$}  >{$}c<{$} |@{\extracolsep{\fill}}>{$}c<{$} >{$}c<{$} >{$}c<{$}>{$}c<{$} >{$}c<{$} >{$}c<{$} >{$}c<{$} >{$}c<{$} >{$}c<{$}|}
\hline
\Delta U & \bm{\times~\varepsilon}&  [_{88}^{\rm pp}]_{\bm{\bar{15}}s}^{1} & [_{88}^{\rm pp}]_{\bm{\bar{15}}s}^{2} &[_{88}^{\rm pp}]_{\bm{\bar{15}}s}^{3} &[_{88}^{\rm pp}]_{\bm{\bar{15}}s}^{4} & [_{88}^{\rm pp}]_{\bm{\bar{15}}s}^{5} & [_{88}^{\rm pp}]_{\bm{\bar{15}}s}^{6} & [_{88}^{\rm pp}]_{\bm{\bar{15}}s}^{7} & [_{88}^{\rm pp}]_{\bm{\bar{15}}s}^{8} &[_{88}^{\rm pp}]_{\bm{\bar{15}}s}^{9} \\
\hline
\hline
 0 & \sml{D^0\to \pi^- \pi^+} & 0 & 2 \lambda  & -\lambda  & 2 \lambda  & -\frac{\lambda }{2} & \lambda  & -3 \lambda  & -\frac{\lambda }{2} & -\frac{\lambda
   }{2} \\
  -1 & \sml{D^0\to K^- \pi^+} & 0 & -2 & 1 & 1 & \frac{1}{2} & \frac{1}{2} & 0 & -1 & \frac{1}{2} \\
   1 & \sml{D^0\to \pi^- K^+} & 0 & -\lambda^2 & -\lambda^2 & 2 \lambda^2 & \lambda^2 & \lambda^2 & 0 & -\frac{\lambda^2}{2} & -\frac{\lambda^2}{2} \\
0 & \sml{D^0\to K^- K^+} & 0 & \lambda  & \lambda  & \lambda  & -\lambda  & \frac{\lambda }{2} & -3 \lambda  & -\lambda  & \frac{\lambda }{2} \\
\hline
 0 & \sml{D^0\to 2\eta_8} & 2 \lambda  & 0 & \lambda  & 0 & -\frac{\lambda }{2} & -2 \lambda  & -3 \lambda  & -\frac{\lambda }{2} & -\frac{3 \lambda }{2} \\
 1 & \sml{D^0\to K^0 \eta_8} & 0 & \frac{\lambda^2}{\sqrt{6}} & \frac{\lambda^2}{\sqrt{6}} & -\sqrt{\frac{2}{3}} \lambda^2 & -\frac{\lambda^2}{\sqrt{6}} &
   -\sqrt{\frac{2}{3}} \lambda^2 & 0 & -\frac{\lambda^2}{2 \sqrt{6}} & -\frac{5 \lambda^2}{2 \sqrt{6}} \\
 -1 & \sml{D^0\to \eta_8 \bar{K}^0} & 0 & \sqrt{\frac{2}{3}} & -\frac{1}{\sqrt{6}} & -\frac{1}{\sqrt{6}} & \frac{5}{2 \sqrt{6}} & \sqrt{\frac{2}{3}} & 0 &
   \frac{1}{2 \sqrt{6}} & \frac{5}{2 \sqrt{6}} \\
 0 & \sml{D^0\to K^0 \bar{K}^0} & 3 \lambda  & 0 & 0 & 0 & \frac{3 \lambda }{2} & 0 & -3 \lambda  & 0 & \frac{3 \lambda }{2}\\
\hline
0 & \sml{D^0\to 2\pi_0} & 0 & 2 \lambda  & -\lambda  & 2 \lambda  & -\frac{\lambda }{2} & -\lambda  & -3 \lambda  & \frac{\lambda }{2} & -\frac{\lambda }{2}
   \\
 0 & \sml{D^0\to \pi_0 \eta_8} & -\sqrt{3} \lambda  & \frac{\lambda }{\sqrt{3}} & \frac{\lambda }{\sqrt{3}} & \frac{\lambda }{\sqrt{3}} & \frac{2 \lambda
   }{\sqrt{3}} & \frac{5 \lambda }{2 \sqrt{3}} & 0 & -\frac{\lambda }{2 \sqrt{3}} & \frac{\lambda }{2 \sqrt{3}} \\
 1 & \sml{D^0\to K^0 \pi_0} & 0 & \frac{\lambda^2}{\sqrt{2}} & \frac{\lambda^2}{\sqrt{2}} & -\sqrt{2} \lambda^2 & \sqrt{2} \lambda^2 & \frac{\lambda
  ^2}{\sqrt{2}} & 0 & -\frac{\lambda^2}{2 \sqrt{2}} & \frac{\lambda^2}{2 \sqrt{2}} \\
 -1 & \sml{D^0\to \pi_0 \bar{K}^0} & 0 & \sqrt{2} & -\frac{1}{\sqrt{2}} & -\frac{1}{\sqrt{2}} & -\frac{1}{2 \sqrt{2}} & -\frac{1}{\sqrt{2}} & 0 & \frac{1}{2
   \sqrt{2}} & -\frac{1}{2 \sqrt{2}} \\
  \hline
 0 & \sml{D^+\to \pi_0 \pi^+} & 0 & 0 & 0 & 0 & 0 & -\sqrt{2} \lambda  & 0 & \frac{\lambda }{\sqrt{2}} & 0 \\
 1 & \sml{D^+\to \pi_0 K^+} & 0 & -\frac{\lambda^2}{\sqrt{2}} & -\frac{\lambda^2}{\sqrt{2}} & \sqrt{2} \lambda^2 & -\sqrt{2} \lambda^2 & -\frac{\lambda
  ^2}{\sqrt{2}} & 0 & \frac{\lambda^2}{2 \sqrt{2}} & -\frac{\lambda^2}{2 \sqrt{2}} \\
 0 & \sml{D_s^+\to \pi_0 K^+} & -\frac{3 \lambda }{\sqrt{2}} & \frac{\lambda }{\sqrt{2}} & -\sqrt{2} \lambda  & -\sqrt{2} \lambda  & \sqrt{2}
   \lambda  & \frac{\lambda }{2 \sqrt{2}} & 0 & \frac{\lambda }{2 \sqrt{2}} & \frac{\lambda }{2 \sqrt{2}} \\
 0 & \sml{D^+\to \eta_8 \pi^+} & -\sqrt{6} \lambda  & -\sqrt{\frac{2}{3}} \lambda  & -\sqrt{\frac{2}{3}} \lambda  & -\sqrt{\frac{2}{3}} \lambda  &
   \sqrt{\frac{2}{3}} \lambda  & \frac{\lambda }{\sqrt{6}} & 0 & -\sqrt{\frac{2}{3}} \lambda  & -\frac{\lambda }{\sqrt{6}} \\
 -1 & \sml{D_s^+\to \eta_8 \pi^+} & 0 & \sqrt{\frac{2}{3}} & -2 \sqrt{\frac{2}{3}} & \sqrt{\frac{2}{3}} & -\sqrt{\frac{2}{3}} &
   -\frac{1}{\sqrt{6}} & 0 & \sqrt{\frac{2}{3}} & \frac{1}{\sqrt{6}} \\
 1 & \sml{D^+\to K^0 \pi^+} & 0 & -\lambda^2 & -\lambda^2 & 2 \lambda^2 & \lambda^2 & \lambda^2 & 0 & -\frac{\lambda^2}{2} & -\frac{\lambda^2}{2} \\
 0 & \sml{D_s^+\to K^0 \pi^+} & -3 \lambda  & \lambda  & -2 \lambda  & -2 \lambda  & -\lambda  & -\frac{\lambda }{2} & 0 & -\frac{\lambda }{2} &
   \frac{\lambda }{2} \\
 -1 & \sml{D^+\to \bar{K}^0 \pi^+} & 0 & 0 & 0 & 0 & 0 & -\frac{1}{2} & 0 & -\frac{1}{2} & 0 \\
 1 & \sml{D^+\to \eta_8 K^+} & 0 & \frac{\lambda^2}{\sqrt{6}} & \frac{\lambda^2}{\sqrt{6}} & -\sqrt{\frac{2}{3}} \lambda^2 & -\frac{\lambda^2}{\sqrt{6}} &
   -\sqrt{\frac{2}{3}} \lambda^2 & 0 & -\frac{\lambda^2}{2 \sqrt{6}} & -\frac{5 \lambda^2}{2 \sqrt{6}} \\
 0 & \sml{D_s^+\to \eta_8 K^+} & \sqrt{\frac{3}{2}} \lambda  & -\frac{\lambda }{\sqrt{6}} & \sqrt{\frac{2}{3}} \lambda  & \sqrt{\frac{2}{3}}
   \lambda  & \frac{\lambda }{\sqrt{6}} & -\frac{5 \lambda }{2 \sqrt{6}} & 0 & \frac{\lambda }{2 \sqrt{6}} & \frac{5 \lambda }{2 \sqrt{6}} \\
 1 & \sml{D_s^+\to K^0 K^+} & 0 & 0 & 0 & 0 & 0 & -\lambda^2 & 0 & -\lambda^2 & 0 \\
  0 & \sml{D^+\to \bar{K}^0 K^+} & -3 \lambda  & -\lambda  & -\lambda  & -\lambda  & -\frac{\lambda }{2} & \frac{\lambda }{2} & 0 & -\lambda  & \lambda  \\
 -1 & \sml{D_s^+\to \bar{K}^0 K^+} & 0 & 1 & -2 & 1 & \frac{1}{2} & \frac{1}{2} & 0 & \frac{1}{2} & -1 \\
\hline
\end{tabular*}
}
\caption{$D \to P_8P_8$ invariants at first order in spurion, generated by $\bm{\bar{15}}$.}
\label{tab:PP88S15}
\end{table}

\clearpage

\section{\texorpdfstring{$D \to PV$}{DPV} Invariants}
\label{app:PV}
\subsection{\texorpdfstring{$\mathcal{O}(1)$}{O1} Invariants}
In the PV case, there are significantly more invariants for each irrep, since the invariants that are antisymmetric in the out-states are no longer zero. There are similarly twice as many amplitudes,  simply because there are two ways to replace a P with a V in each. The $\mathcal{O}(1)$ invariants are shown in Tables \ref{tab:PV111881}, \ref{tab:PV88D0} and \ref{tab:PV88DP} below, for the $P_1V_1$, $P_1V_8$, $V_1P_8$ and $P_8V_8$ amplitudes. Similarly to Appendix \ref{app:PP}, each invariant is labelled by $[_{xy}^{\rm pv}]^k_R$ as in eq. (\ref{eqn:AIL}), where $x,y$=$1,8$ labels the P and V representation respectively, $R$ is the $H$ representation as appropriate, and $k$ indicates the $k$th such invariant.

\begin{table}[ht]
\begin{tabular*}{\linewidth}{|>{$}c<{$}  >{$}c<{$} |@{\extracolsep{\fill}} >{$}c<{$}  >{$}c<{$} >{$}c<{$}  >{$}c<{$} >{$}c<{$}  >{$}c<{$}  >{$}c<{$}|}
\hline
 \Delta U & \mbox{Ampl.}   & [_{11}^{\rm pv}]_{\bm{\bar{3}}}  & [_{18}^{\rm pv}]_{\bm{\bar{3}}}  & [_{18}^{\rm pv}]_{\bm{6}}  & [_{18}^{\rm pv}]_{\bm{\bar{15}}} & [_{81}^{\rm pv}]_{\bm{\bar{3}}}  & [_{81}^{\rm pv}]_{\bm{6}} & [_{81}^{\rm pv}]_{\bm{\bar{15}}}\\   
 \hline
 \hline
  0 & \sml{D^0\to \eta_1 \phi_1} & \frac{\Delta }{8} & 0 & 0 & 0 & 0 & 0 & 0 \\
 0 & \sml{D^0\to \eta_1 \omega_8} & 0 & \frac{\Delta }{8 \sqrt{6}} & -\frac{\lambda}{2} \sqrt{\frac{3}{2}}  & -\frac{\lambda}{2} \sqrt{\frac{3}{2}}  & 0 & 0
   & 0 \\
 0 & \sml{D^0\to \eta_1 \rho_0} & 0 & \frac{\Delta }{8 \sqrt{2}} & \frac{\lambda }{2 \sqrt{2}} & \frac{\lambda }{2 \sqrt{2}} & 0 & 0 & 0 \\
 1 & \sml{D^0\to \eta_1 K^{*0}} & 0 & 0 & -\frac{\lambda^2}{2} & -\frac{\lambda^2}{2} & 0 & 0 & 0 \\
 -1 & \sml{D^0\to \eta_1 \bar{K}^{*0}} & 0 & 0 & \frac{1}{2} & \frac{1}{2} & 0 & 0 & 0 \\
 0 & \sml{D^0\to \eta_8 \phi_1} & 0 & 0 & 0 & 0 & \frac{\Delta }{8 \sqrt{6}} & -\frac{\lambda}{2} \sqrt{\frac{3}{2}}  & -\frac{ \lambda}{2} \sqrt{\frac{3}{2}}\\
 1 & \sml{D^0\to K^0 \phi_1} & 0 & 0 & 0 & 0 & 0 & -\frac{\lambda^2}{2} & -\frac{\lambda^2}{2} \\
\hline
\end{tabular*}
\caption{$D \to P_1V_1$, $D \to P_1V_8$ and $D \to V_1P_8$ $\mathcal{O}(1)$ invariants. $\Delta = \lambda^5A^2(\rho - i\eta)$.}
\label{tab:PV111881}
\end{table}

\begin{table}[ht]
\resizebox{\linewidth}{!}
{
\begin{tabular*}{1.05\linewidth}{|>{$}c<{$}  >{$}c<{$} |@{\extracolsep{\fill}} >{$}c<{$}  >{$}c<{$} >{$}c<{$}  >{$}c<{$} >{$}c<{$} >{$}c<{$}  >{$}c<{$} >{$}c<{$}  >{$}c<{$} >{$}c<{$}|}
\hline
\Delta U & \mbox{Ampl.}  & [_{88}^{\rm pv}]_{\bm{\bar{3}}}^1 & [_{88}^{\rm pv}]_{\bm{\bar{3}}}^2  &[_{88}^{\rm pv}]_{\bm{\bar{3}}}^3 & [_{88}^{\rm pv}]_{\bm{6}}^1 & [_{88}^{\rm pv}]_{\bm{6}}^2 & [_{88}^{\rm pv}]_{\bm{6}}^3 & [_{88}^{\rm pv}]_{\bm{\bar{15}}}^1 &[_{88}^{\rm pv}]_{\bm{\bar{15}}}^2 & [_{88}^{\rm pv}]_{\bm{\bar{15}}}^3 & [_{88}^{\rm pv}]_{\bm{\bar{15}}}^4\\
\hline\hline
 0 & \sml{D^0\to \eta_8 \omega_8} & -\frac{\Delta }{12} & \frac{\Delta }{8} & \frac{\Delta }{48} & 0 & \frac{\lambda }{4} & -\frac{\lambda }{4} & \frac{\lambda
   }{2} & \frac{\lambda }{4} & \frac{\lambda }{4} & -\frac{\lambda }{4} \\
 0 & \sml{D^0\to \eta_8 \rho_0} & \frac{\Delta }{8 \sqrt{3}} & 0 & \frac{\Delta }{16 \sqrt{3}} & -\frac{\lambda }{2 \sqrt{3}} & -\frac{\lambda }{4 \sqrt{3}} &
   -\frac{\sqrt{3} \lambda }{4} & \frac{\lambda }{2 \sqrt{3}} & \frac{5 \lambda }{4 \sqrt{3}} & \frac{\lambda }{4 \sqrt{3}} & \frac{\lambda }{4 \sqrt{3}} \\
 1 & \sml{D^0\to \eta_8 K^{*0}} & 0 & 0 & 0 & \frac{\lambda^2}{\sqrt{6}} & \frac{\lambda^2}{2 \sqrt{6}} & 0 & \frac{\lambda^2}{2 \sqrt{6}} & \frac{\lambda
  ^2}{2 \sqrt{6}} & -\frac{\lambda^2}{2 \sqrt{6}} & -\frac{\lambda^2}{2 \sqrt{6}} \\
 -1 & \sml{D^0\to \eta_8 \bar{K}^{*0}} & 0 & 0 & 0 & \frac{1}{2 \sqrt{6}} & -\frac{1}{2 \sqrt{6}} & 0 & -\frac{1}{2 \sqrt{6}} & \frac{1}{\sqrt{6}} &
   -\frac{1}{\sqrt{6}} & \frac{1}{2 \sqrt{6}} \\
 1 & \sml{D^0\to K^0 \omega_8} & 0 & 0 & 0 & -\frac{\lambda^2}{\sqrt{6}} & 0 & -\frac{\lambda^2}{2 \sqrt{6}} & \frac{\lambda^2}{2 \sqrt{6}} & 0 &
   \frac{\lambda^2}{\sqrt{6}} & 0 \\
 1 & \sml{D^0\to K^0 \rho_0} & 0 & 0 & 0 & 0 & 0 & -\frac{\lambda^2}{2 \sqrt{2}} & \frac{\lambda^2}{2 \sqrt{2}} & \frac{\lambda^2}{\sqrt{2}} & 0 & 0 \\
\hline
\end{tabular*}
}
\caption{$D^0 \to P_8V_8$ $\mathcal{O}(1)$ invariants. $\Delta = \lambda^5A^2(\rho - i\eta)$.}
\label{tab:PV88D0}
\end{table}
\clearpage 

\begin{table}[ht]
\resizebox{\linewidth}{!}
{
\begin{tabular*}{1.1\linewidth}{|>{$}c<{$}  >{$}c<{$} |@{\extracolsep{\fill}} >{$}c<{$}  >{$}c<{$} >{$}c<{$}  >{$}c<{$} >{$}c<{$} >{$}c<{$}  >{$}c<{$} >{$}c<{$}  >{$}c<{$} >{$}c<{$}|}
\hline
\Delta U &  \mbox{Ampl.}  & [_{88}^{\rm pv}]_{\bm{\bar{3}}}^1 & [_{88}^{\rm pv}]_{\bm{\bar{3}}}^2  &[_{88}^{\rm pv}]_{\bm{\bar{3}}}^3 & [_{88}^{\rm pv}]_{\bm{6}}^1 & [_{88}^{\rm pv}]_{\bm{6}}^2 & [_{88}^{\rm pv}]_{\bm{6}}^3 & [_{88}^{\rm pv}]_{\bm{\bar{15}}}^1 &[_{88}^{\rm pv}]_{\bm{\bar{15}}}^2 & [_{88}^{\rm pv}]_{\bm{\bar{15}}}^3 & [_{88}^{\rm pv}]_{\bm{\bar{15}}}^4\\
\hline\hline
0 & \sml{D^+\to \eta_8 \rho^+} & \frac{\Delta }{4 \sqrt{6}} & 0 & \frac{\Delta }{8 \sqrt{6}} & -\frac{\lambda }{\sqrt{6}} & -\frac{\lambda }{2 \sqrt{6}} &
   -\frac{\lambda}{2} \sqrt{\frac{3}{2}}  & -\frac{\lambda }{\sqrt{6}} & \frac{\lambda }{2 \sqrt{6}} & -\frac{\lambda }{2 \sqrt{6}} & -\frac{\lambda }{2
   \sqrt{6}} \\
 1 & \sml{D^+\to \eta_8 K^{*+}} & 0 & 0 & 0 & -\frac{\lambda^2}{\sqrt{6}} & -\frac{\lambda^2}{2 \sqrt{6}} & 0 & \frac{\lambda^2}{2 \sqrt{6}} &
   \frac{\lambda^2}{2 \sqrt{6}} & -\frac{\lambda^2}{2 \sqrt{6}} & -\frac{\lambda^2}{2 \sqrt{6}} \\
 1 & \sml{D^+\to K^0 \rho^+} & 0 & 0 & 0 & 0 & 0 & -\frac{\lambda^2}{2} & -\frac{\lambda^2}{2} & 0 & 0 & 0 \\
 0 & \sml{D^+\to \pi_0 \rho^+} & 0 & 0 & \frac{\Delta }{8 \sqrt{2}} & 0 & \frac{\lambda }{2 \sqrt{2}} & \frac{\lambda }{2 \sqrt{2}} & 0 & \frac{\lambda }{2
   \sqrt{2}} & -\frac{\lambda }{2 \sqrt{2}} & \frac{\lambda }{2 \sqrt{2}} \\
 1 & \sml{D^+\to \pi_0 K^{*+}} & 0 & 0 & 0 & 0 & \frac{\lambda^2}{2 \sqrt{2}} & 0 & -\frac{\lambda^2}{2 \sqrt{2}} & \frac{\lambda^2}{2 \sqrt{2}} &
   -\frac{\lambda^2}{2 \sqrt{2}} & \frac{\lambda^2}{2 \sqrt{2}} \\
 -1 & \sml{D^+\to \bar{K}^0 \rho^+} & 0 & 0 & 0 & \frac{1}{2} & \frac{1}{2} & \frac{1}{2} & 0 & 0 & 0 & \frac{1}{2} \\
 0 & \sml{D^+\to \bar{K}^0 K^{*+}} & \frac{\Delta }{8} & 0 & 0 & \frac{\lambda }{2} & \frac{\lambda }{2} & 0 & -\frac{\lambda }{2} & 0 & 0 &
   \frac{\lambda }{2} \\
\hline
\end{tabular*}
}
\caption{$D^+ \to P_8V_8$ and $D^+_s \to P_8V_8$  $\mathcal{O}(1)$ invariants. $\Delta = \lambda^5A^2(\rho - i\eta)$.}
\label{tab:PV88DP}
\end{table}
\clearpage 

\subsection{Spurionic \texorpdfstring{$\mathcal{O}(\varepsilon)$}{OE} Invariants}
We finally present the invariants produced by $Hm_s$ for $D \to PV$, shown in Tables \ref{tab:PV1881S} to \ref{tab:PV88S15DP} following. Each invariant is labelled as in the previous section, but with an extra $s$ subscript. 

\begin{table}[ht]
\begin{center}
\begin{tabular*}{0.7\linewidth}{|>{$}c<{$}  >{$}c<{$} |@{\extracolsep{\fill}} >{$}c<{$}  >{$}c<{$} >{$}c<{$} |}
\hline
\Delta U & \bm{\times ~\varepsilon} & [_{11}^{\rm pv}]_{\bm{\bar{3}}s}  & [_{11}^{\rm pv}]_{\bm{6}s} & [_{11}^{\rm pv}]_{\bm{\bar{15}}s}\\
\hline
\hline
0 & \sml{D^0\to \phi_1  \eta _1} & \frac{\Delta }{8}  & -\frac{3   \lambda }{2}  & -\frac{3  \lambda }{2}\\
\hline
\end{tabular*}
\end{center}
\vspace{1cm}

\resizebox{\linewidth}{!}
{
\begin{tabular*}{1.08\linewidth}{|>{$}c<{$}  >{$}c<{$} |@{\extracolsep{\fill}} >{$}c<{$}  >{$}c<{$} >{$}c<{$}  >{$}c<{$} >{$}c<{$} >{$}c<{$}  >{$}c<{$} >{$}c<{$}  >{$}c<{$} >{$}c<{$}|}
\hline
 \Delta U &  \bm{\times ~\varepsilon}& [_{18}^{\rm pv}]_{\bm{\bar{3}}s}^1 & [_{18}^{\rm pv}]_{\bm{\bar{3}}s}^2  &[_{18}^{\rm pv}]_{\bm{\bar{3}}s}^3 & [_{18}^{\rm pv}]_{\bm{6}s}^1 & [_{18}^{\rm pv}]_{\bm{6}s}^2 & [_{18}^{\rm pv}]_{\bm{6}s}^3 & [_{18}^{\rm pv}]_{\bm{\bar{15}}s}^1 &[_{18}^{\rm pv}]_{\bm{\bar{15}}s}^2 & [_{18}^{\rm pv}]_{\bm{\bar{15}}s}^3 & [_{18}^{\rm pv}]_{\bm{\bar{15}}s}^4\\
\hline
\hline
 0 & \sml{D^0\to \eta_1 \omega_8} & -\frac{\Delta }{2 \sqrt{6}} & \frac{\Delta }{8 \sqrt{6}} & \frac{\Delta }{8 \sqrt{6}} & \frac{ \lambda}{2} \sqrt{\frac{3}{2}}
    & 0 & \frac{\lambda}{2} \sqrt{\frac{3}{2}}  & \frac{\lambda}{2} \sqrt{\frac{3}{2}}  & -\sqrt{\frac{3}{2}} \lambda  & -\frac{\lambda}{2}
   \sqrt{\frac{3}{2}}   & \frac{\lambda}{2} \sqrt{\frac{3}{2}}  \\
 0 & \sml{D^0\to \eta_1 \rho_0} & \frac{\Delta }{4 \sqrt{2}} & \frac{\Delta }{8 \sqrt{2}} & \frac{\Delta }{8 \sqrt{2}} & \frac{\lambda }{2 \sqrt{2}} &
   -\frac{\lambda }{\sqrt{2}} & -\frac{\lambda }{2 \sqrt{2}} & -\frac{3 \lambda }{2 \sqrt{2}} & -\frac{\lambda }{\sqrt{2}} & -\frac{3 \lambda }{2 \sqrt{2}} &
   \frac{\lambda }{2 \sqrt{2}} \\
 1 & \sml{D^0\to \eta_1 K^{*0}} & 0 & 0 & 0 & \lambda^2 & \lambda^2 & \frac{\lambda^2}{2} & 0 & -\frac{\lambda^2}{2} & 0 & -\frac{\lambda^2}{2} \\
 -1 & \sml{D^0\to \eta_1 \bar{K}^{*0}} & 0 & 0 & 0 & \frac{1}{2} & \frac{1}{2} & -\frac{1}{2} & -\frac{3}{2} & \frac{1}{2} & 0 & -1 \\
 0 & \sml{D^+\to \eta_1 \rho^+} & \frac{\Delta }{4} & \frac{\Delta }{8} & \frac{\Delta }{8} & \frac{\lambda }{2} & -\lambda  & -\frac{\lambda }{2} & -\frac{3
   \lambda }{2} & \lambda  & -\frac{3 \lambda }{2} & -\frac{\lambda }{2} \\
 1 & \sml{D^+\to \eta_1 K^{*+}} & 0 & 0 & 0 & -\lambda^2 & -\lambda^2 & -\frac{\lambda^2}{2} & 0 & -\frac{\lambda^2}{2} & 0 & -\frac{\lambda
  ^2}{2} \\
\hline
\end{tabular*}
}
\vspace{1 cm}

\resizebox{\linewidth}{!}
{
\begin{tabular*}{1.08\linewidth}{|>{$}c<{$}  >{$}c<{$} |@{\extracolsep{\fill}} >{$}c<{$}  >{$}c<{$} >{$}c<{$}  >{$}c<{$} >{$}c<{$} >{$}c<{$}  >{$}c<{$} >{$}c<{$}  >{$}c<{$} >{$}c<{$}|}
\hline
 \Delta U &   \bm{\times ~ \varepsilon}& [_{81}^{\rm pv}]_{\bm{\bar{3}}s}^1 & [_{81}^{\rm pv}]_{\bm{\bar{3}}s}^2  &[_{81}^{\rm pv}]_{\bm{\bar{3}}s}^3 & [_{81}^{\rm pv}]_{\bm{6}s}^1 & [_{81}^{\rm pv}]_{\bm{6}s}^2 & [_{81}^{\rm pv}]_{\bm{6}s}^3 & [_{81}^{\rm pv}]_{\bm{\bar{15}}s}^1 &[_{81}^{\rm pv}]_{\bm{\bar{15}}s}^2 & [_{81}^{\rm pv}]_{\bm{\bar{15}}s}^3 & [_{81}^{\rm pv}]_{\bm{\bar{15}}s}^4\\
\hline
\hline
 0 & \sml{D^0\to \eta_8 \phi_1} & -\frac{\Delta }{2 \sqrt{6}} & \frac{\Delta }{8 \sqrt{6}} & \frac{\Delta }{8 \sqrt{6}} & \frac{\lambda}{2} \sqrt{\frac{3}{2}}
    & 0 & \frac{\lambda}{2} \sqrt{\frac{3}{2}}  & \frac{\lambda}{2} \sqrt{\frac{3}{2}}  & -\sqrt{\frac{3}{2}} \lambda  & -\frac{\lambda}{2} \sqrt{\frac{3}{2}}
  & \frac{\lambda}{2} \sqrt{\frac{3}{2}}  \\
 1 & \sml{D^0\to K^0 \phi_1} & 0 & 0 & 0 & \lambda^2 & \lambda^2 & \frac{\lambda^2}{2} & 0 & -\frac{\lambda^2}{2} & 0 & -\frac{\lambda^2}{2} \\
 0 & \sml{D^0\to \pi_0 \phi_1} & \frac{\Delta }{4 \sqrt{2}} & \frac{\Delta }{8 \sqrt{2}} & \frac{\Delta }{8 \sqrt{2}} & \frac{\lambda }{2 \sqrt{2}} &
   -\frac{\lambda }{\sqrt{2}} & -\frac{\lambda }{2 \sqrt{2}} & -\frac{3 \lambda }{2 \sqrt{2}} & -\frac{\lambda }{\sqrt{2}} & -\frac{3 \lambda }{2 \sqrt{2}} &
   \frac{\lambda }{2 \sqrt{2}} \\
 -1 & \sml{D^0\to \phi_1 \bar{K}^0} & 0 & 0 & 0 & \frac{1}{2} & \frac{1}{2} & -\frac{1}{2} & -\frac{3}{2} & \frac{1}{2} & 0 & -1 \\
 1 & \sml{D^+\to \phi_1 K^+} & 0 & 0 & 0 & -\lambda^2 & -\lambda^2 & -\frac{\lambda^2}{2} & 0 & -\frac{\lambda^2}{2} & 0 & -\frac{\lambda^2}{2} \\
 0 & \sml{D^+\to \phi_1 \pi^+} & \frac{\Delta }{4} & \frac{\Delta }{8} & \frac{\Delta }{8} & \frac{\lambda }{2} & -\lambda  & -\frac{\lambda }{2} & -\frac{3
   \lambda }{2} & \lambda  & -\frac{3 \lambda }{2} & -\frac{\lambda }{2} \\
\hline
\end{tabular*}
}

\caption{$D \to P_1V_1$, $D \to P_1V_8$ and $D \to P_8V_1$ $\mathcal{O}(\varepsilon)$ invariants. $\Delta = \lambda^5A^2(\rho - i\eta)$.}
\label{tab:PV1881S}
\end{table}

\clearpage

\begin{table}[ht]
\begin{center}
\rotatebox{270}{
\resizebox{\linewidth}{!}
{
\begin{tabular*}{1.1\linewidth}{|>{$}c<{$}  >{$}c<{$} |@{\extracolsep{\fill}} >{$}c<{$}  >{$}c<{$} >{$}c<{$}  >{$}c<{$} >{$}c<{$} >{$}c<{$}  >{$}c<{$} >{$}c<{$}  >{$}c<{$} >{$}c<{$}|}
\hline
\Delta U & \bm{ \times \varepsilon \Delta}   &[_{88}^{\rm pv}]^1_{\bm{\bar{3}}s} &[_{88}^{\rm pv}]^2_{\bm{\bar{3}}s} & [_{88}^{\rm pv}]^3_{\bm{\bar{3}}s} & [_{88}^{\rm pv}]^4_{\bm{\bar{3}}s} & [_{88}^{\rm pv}]^5_{\bm{\bar{3}}s}& [_{88}^{\rm pv}]^6_{\bm{\bar{3}}s} & [_{88}^{\rm pv}]^7_{\bm{\bar{3}}s} & [_{88}^{\rm pv}]^8_{\bm{\bar{3}}s} & [_{88}^{\rm pv}]^9_{\bm{\bar{3}}s} & [_{88}^{\rm pv}]^{10}_{\bm{\bar{3}}s}\\
\hline
\hline
 0 & \sml{D^0\to \eta_8 \omega_8} & -\frac{1}{12} & -\frac{1}{24} & -\frac{1}{12} & -\frac{1}{4} & -\frac{1}{12} & -\frac{1}{24} & \frac{1}{8} & \frac{1}{48} &
   \frac{1}{48} & \frac{1}{8} \\
 0 & \sml{D^0\to \eta_8 \rho_0} & \frac{1}{8 \sqrt{3}} & -\frac{1}{8 \sqrt{3}} & \frac{1}{8 \sqrt{3}} & 0 & -\frac{1}{4 \sqrt{3}} & -\frac{1}{8 \sqrt{3}} & 0 &
   \frac{1}{16 \sqrt{3}} & \frac{1}{16 \sqrt{3}} & 0 \\
 0 & \sml{D^0\to K^0 \bar{K}^{*0}} & -\frac{1}{8} & \frac{1}{8} & -\frac{1}{8} & -\frac{1}{8} & 0 & -\frac{1}{4} & \frac{1}{8} & 0 & 0 & 0 \\
 0 & \sml{D^0\to \pi_0 \omega_8} & \frac{1}{8 \sqrt{3}} & \frac{1}{4 \sqrt{3}} & \frac{1}{8 \sqrt{3}} & 0 & \frac{1}{8 \sqrt{3}} & \frac{1}{4 \sqrt{3}} & 0 &
   \frac{1}{16 \sqrt{3}} & \frac{1}{16 \sqrt{3}} & \frac{\sqrt{3}}{8} \\
 0 & \sml{D^0\to \pi_0 \rho_0} & 0 & 0 & 0 & \frac{1}{4} & \frac{1}{8} & 0 & \frac{1}{8} & \frac{1}{16} & \frac{1}{16} & 0 \\
 0 & \sml{D^0\to \bar{K}^0 K^{*0}} & -\frac{1}{8} & -\frac{1}{4} & -\frac{1}{8} & -\frac{1}{8} & 0 & \frac{1}{8} & \frac{1}{8} & 0 & 0 & 0 \\
 0 & \sml{D^0\to K^- K^{*+}} & 0 & 0 & 0 & -\frac{1}{8} & 0 & 0 & \frac{1}{8} & 0 & 0 & 0 \\
\hline
\end{tabular*}
}}
\end{center}
\caption{$D \to P_8V_8$ $\mathcal{O}(\varepsilon)$ invariants generated by $\bm{\bar{3}}$. $\Delta = \lambda^5A^2(\rho - i\eta)$.}
\label{tab:PV88S3}
\end{table}

\clearpage

\begin{table}[ht]
\begin{center}
\rotatebox{270}{
\resizebox{1.2\linewidth}{!}{
\begin{tabular*}{1.3\linewidth}{|>{$}c<{$}  >{$}c<{$} |@{\extracolsep{\fill}} >{$}c<{$}  >{$}c<{$} >{$}c<{$}  >{$}c<{$} >{$}c<{$} >{$}c<{$}  >{$}c<{$} >{$}c<{$}  >{$}c<{$} >{$}c<{$}>{$}c<{$} >{$}c<{$}|}
\hline
\Delta U & \bm{\times~\varepsilon}  & [_{88}^{\rm pv}]^1_{\bm{6}s}  & [_{88}^{\rm pv}]^2_{\bm{6}s} & [_{88}^{\rm pv}]^3_{\bm{6}s} & [_{88}^{\rm pv}]^4_{\bm{6}s} & [_{88}^{\rm pv}]^5_{\bm{6}s} & [_{88}^{\rm pv}]^6_{\bm{6}s} & [_{88}^{\rm pv}]^7_{\bm{6}s} & [_{88}^{\rm pv}]^8_{\bm{6}s} & [_{88}^{\rm pv}]^9_{\bm{6}s} & [_{88}^{\rm pv}]^{10}_{\bm{6}s} & [_{88}^{\rm pv}]^{11}_{\bm{6}s} & [_{88}^{\rm pv}]^{12}_{\bm{6}s} \\
\hline
\hline
  0   &   \sml{D^0\to \omega_8  \eta_8}   &   0   &   0   &   \frac{\lambda }{2}   &   -\frac{\lambda }{2}   &   -\frac{\lambda }{4}   &   -\frac{\lambda }{4}   &   \frac{\lambda }{4}   &   0   &   0   &   0
     &   -\frac{3 \lambda }{2}   &   -\frac{\lambda }{2}   \\  
 0   &   \sml{D^0\to \eta_8 \rho_0}   &   \frac{\lambda }{2 \sqrt{3}}   &   -\frac{\lambda }{2 \sqrt{3}}   &   -\frac{\lambda }{2 \sqrt{3}}   &   \frac{\lambda }{\sqrt{3}}   &  
   -\frac{\lambda }{4 \sqrt{3}}   &   -\frac{\sqrt{3} \lambda }{4}   &   \frac{\sqrt{3} \lambda }{4}   &   0   &   -\frac{\lambda }{2 \sqrt{3}}   &   \frac{\lambda }{2 \sqrt{3}}   &  
   0   &   \frac{\lambda }{2 \sqrt{3}}   \\  
 1   &   \sml{D^0\to \eta_8 K^{*0}}   &   \frac{\lambda^2}{2 \sqrt{6}}   &   -\frac{\lambda^2}{2 \sqrt{6}}   &   -\sqrt{\frac{2}{3}} \lambda^2   &   -\sqrt{\frac{2}{3}} \lambda
  ^2   &   -\frac{\lambda^2}{\sqrt{6}}   &   0   &   0   &   0   &   \frac{\lambda^2}{\sqrt{6}}   &   -\frac{\lambda^2}{\sqrt{6}}   &   0   &   -\frac{\lambda^2}{\sqrt{6}}   \\  
 -1   &   \sml{D^0\to \eta_8 \bar{K}^{*0}}   &   \frac{1}{\sqrt{6}}   &   -\frac{1}{\sqrt{6}}   &   \frac{1}{2 \sqrt{6}}   &   \frac{1}{2 \sqrt{6}}   &   -\frac{1}{2 \sqrt{6}}   &   0   &   0
     &   0   &   \frac{1}{2 \sqrt{6}}   &   -\frac{1}{2 \sqrt{6}}   &   0   &   \frac{1}{\sqrt{6}}   \\
\hline
\end{tabular*}
}}
\end{center}
\caption{$D^0 \to P_8V_8$ $\mathcal{O}(\varepsilon)$ invariants generated by $\bm{6}$.}
\label{tab:PV88S6D0}
\end{table}

\clearpage

\begin{table}[ht]
\begin{center}
\rotatebox{270}{
\resizebox{1.2\linewidth}{!}{
\begin{tabular*}{1.3\linewidth}{|>{$}c<{$}  >{$}c<{$} |@{\extracolsep{\fill}} >{$}c<{$}  >{$}c<{$} >{$}c<{$}  >{$}c<{$} >{$}c<{$} >{$}c<{$}  >{$}c<{$} >{$}c<{$}  >{$}c<{$} >{$}c<{$}>{$}c<{$} >{$}c<{$}|}
\hline
\Delta U & \bm{\times~\varepsilon}  & [_{88}^{\rm pv}]^1_{\bm{6}s}  & [_{88}^{\rm pv}]^2_{\bm{6}s} & [_{88}^{\rm pv}]^3_{\bm{6}s} & [_{88}^{\rm pv}]^4_{\bm{6}s} & [_{88}^{\rm pv}]^5_{\bm{6}s} & [_{88}^{\rm pv}]^6_{\bm{6}s} & [_{88}^{\rm pv}]^7_{\bm{6}s} & [_{88}^{\rm pv}]^8_{\bm{6}s} & [_{88}^{\rm pv}]^9_{\bm{6}s} & [_{88}^{\rm pv}]^{10}_{\bm{6}s} & [_{88}^{\rm pv}]^{11}_{\bm{6}s} & [_{88}^{\rm pv}]^{12}_{\bm{6}s} \\
\hline
\hline
  0   &   \sml{D^+\to \eta_8 \rho^+}   &   \frac{\lambda }{\sqrt{6}}   &   -\frac{\lambda }{\sqrt{6}}   &   -\frac{\lambda }{\sqrt{6}}   &   \sqrt{\frac{2}{3}} \lambda    &  
   -\frac{\lambda }{2 \sqrt{6}}   &   -\frac{ \lambda}{2} \sqrt{\frac{3}{2}}     &   \frac{ \lambda}{2} \sqrt{\frac{3}{2}}    &   0   &   -\frac{\lambda }{\sqrt{6}}   &  
   \frac{\lambda }{\sqrt{6}}   &   0   &   \frac{\lambda }{\sqrt{6}}   \\  
 1   &   \sml{D^+\to \eta_8 K^{*+}}   &   -\frac{\lambda^2}{2 \sqrt{6}}   &   \frac{\lambda^2}{2 \sqrt{6}}   &   \sqrt{\frac{2}{3}} \lambda^2   &   \sqrt{\frac{2}{3}}
   \lambda^2   &   \frac{\lambda^2}{\sqrt{6}}   &   0   &   0   &   0   &   -\frac{\lambda^2}{\sqrt{6}}   &   \frac{\lambda^2}{\sqrt{6}}   &   0   &   \frac{\lambda^2}{\sqrt{6}}   \\  
 1   &   \sml{D^+\to K^0 \rho^+}   &   \frac{\lambda^2}{2}   &   0   &   0   &   0   &   0   &   -\frac{\lambda^2}{2}   &   \lambda^2   &   \lambda^2   &   0   &   0   &   0   &   0   \\  
 0   &   \sml{D^+\to \pi_0 \rho^+}   &   0   &   -\frac{\lambda }{\sqrt{2}}   &   0   &   0   &   \frac{\lambda }{2 \sqrt{2}}   &   \frac{\lambda }{2 \sqrt{2}}   &   \frac{\lambda }{2
   \sqrt{2}}   &   -\frac{\lambda }{\sqrt{2}}   &   0   &   -\frac{\lambda }{\sqrt{2}}   &   0   &   0   \\
\hline
\end{tabular*}
}}
\end{center}
\caption{$D^+ \to P_8V_8$ and $D^+_s \to P_8V_8$ $\mathcal{O}(\varepsilon)$ invariants generated by $\bm{6}$.}
\label{tab:PV88S6DP}
\end{table}

\clearpage

\begin{table}[ht]
\begin{center}
\rotatebox{270}{
\resizebox{1.25\linewidth}{0.35\linewidth}
{
\begin{tabular*}{2\linewidth}{|>{$}c<{$}  >{$}c<{$} |@{\extracolsep{\fill}} >{$}c<{$}  >{$}c<{$} >{$}c<{$}  >{$}c<{$} >{$}c<{$} >{$}c<{$}  >{$}c<{$} >{$}c<{$}  >{$}c<{$} >{$}c<{$}>{$}c<{$} >{$}c<{$} >{$}c<{$}  >{$}c<{$} >{$}c<{$}  >{$}c<{$} >{$}c<{$} >{$}c<{$}|}
\hline
\Delta U & \bm{\times~\varepsilon}  & [_{88}^{\rm pv}]^1_ {\bm{\bar{15}}s}  & [_{88}^{\rm pv}]^2_ {\bm{\bar{15}}s} & [_{88}^{\rm pv}]^3_ {\bm{\bar{15}}s} & [_{88}^{\rm pv}]^4_ {\bm{\bar{15}}s} & [_{88}^{\rm pv}]^5_ {\bm{\bar{15}}s} & [_{88}^{\rm pv}]^6_ {\bm{\bar{15}}s} & [_{88}^{\rm pv}]^7_ {\bm{\bar{15}}s} & [_{88}^{\rm pv}]^8_ {\bm{\bar{15}}s} & [_{88}^{\rm pv}]^9_ {\bm{\bar{15}}s} & [_{88}^{\rm pv}]^{10}_ {\bm{\bar{15}}s} & [_{88}^{\rm pv}]^{11}_ {\bm{\bar{15}}s} & [_{88}^{\rm pv}]^{12}_ {\bm{\bar{15}}s} & [_{88}^{\rm pv}]^{13}_ {\bm{\bar{15}}s} & [_{88}^{\rm pv}]^{14}_ {\bm{\bar{15}}s} & [_{88}^{\rm pv}]^{15}_ {\bm{\bar{15}}s} & [_{88}^{\rm pv}]^{16}_ {\bm{\bar{15}}s} & [_{88}^{\rm pv}]^{17}_ {\bm{\bar{15}}s} & [_{88}^{\rm pv}]^{18}_ {\bm{\bar{15}}s} \\
\hline
\hline
 0  &  \sml{D^0\to \omega_8  \eta_8}  &  \lambda   &  \frac{\lambda }{2}  &  0  &  \frac{\lambda }{2}  &  \frac{\lambda }{4}  &  0  &  0  &  0  &  0  &  -\frac{\lambda }{4}  & 
   -\frac{\lambda }{2}  &  -\lambda   &  0  &  -\frac{3 \lambda }{2}  &  -\frac{\lambda }{4}  &  -\frac{3 \lambda }{4}  &  -\frac{\lambda }{4}  &  -\frac{3 \lambda }{4}  \\ 
 0  &  \sml{D^0\to \eta_8 \rho_0}  &  -\frac{\sqrt{3} \lambda }{2}  &  0  &  \frac{\lambda }{2 \sqrt{3}}  &  \frac{\lambda }{2 \sqrt{3}}  &  \frac{\sqrt{3} \lambda }{4}  & 
   \frac{\lambda }{2 \sqrt{3}}  &  \frac{\lambda }{2 \sqrt{3}}  &  \frac{\lambda }{\sqrt{3}}  &  \frac{\lambda }{2 \sqrt{3}}  &  \frac{5 \lambda }{4 \sqrt{3}}  & 
   \frac{\lambda }{2 \sqrt{3}}  &  \frac{\lambda }{\sqrt{3}}  &  -\frac{\lambda }{\sqrt{3}}  &  0  &  -\frac{\sqrt{3} \lambda }{4}  &  \frac{\lambda }{4 \sqrt{3}}  & 
   \frac{\lambda }{4 \sqrt{3}}  &  \frac{\lambda }{4 \sqrt{3}}  \\
\hline
\end{tabular*}
}
}
\end{center}
\caption{$D^0 \to P_8V_8$ $\mathcal{O}(\varepsilon)$ invariants generated by $\bm{\bar{15}}$.}
\label{tab:PV88S15D0}
\end{table} 

\clearpage

\begin{table}[ht]
\begin{center}
\rotatebox{270}{
\resizebox{1.25\linewidth}{0.45\linewidth}
{
\begin{tabular*}{2\linewidth}{|>{$}c<{$}  >{$}c<{$} |@{\extracolsep{\fill}} >{$}c<{$}  >{$}c<{$} >{$}c<{$}  >{$}c<{$} >{$}c<{$} >{$}c<{$}  >{$}c<{$} >{$}c<{$}  >{$}c<{$} >{$}c<{$}>{$}c<{$} >{$}c<{$} >{$}c<{$}  >{$}c<{$} >{$}c<{$}  >{$}c<{$} >{$}c<{$} >{$}c<{$}|}
\hline
\Delta U & \bm{\times~\varepsilon}  & [_{88}^{\rm pv}]^1_ {\bm{\bar{15}}s}  & [_{88}^{\rm pv}]^2_ {\bm{\bar{15}}s} & [_{88}^{\rm pv}]^3_ {\bm{\bar{15}}s} & [_{88}^{\rm pv}]^4_ {\bm{\bar{15}}s} & [_{88}^{\rm pv}]^5_ {\bm{\bar{15}}s} & [_{88}^{\rm pv}]^6_ {\bm{\bar{15}}s} & [_{88}^{\rm pv}]^7_ {\bm{\bar{15}}s} & [_{88}^{\rm pv}]^8_ {\bm{\bar{15}}s} & [_{88}^{\rm pv}]^9_ {\bm{\bar{15}}s} & [_{88}^{\rm pv}]^{10}_ {\bm{\bar{15}}s} & [_{88}^{\rm pv}]^{11}_ {\bm{\bar{15}}s} & [_{88}^{\rm pv}]^{12}_ {\bm{\bar{15}}s} & [_{88}^{\rm pv}]^{13}_ {\bm{\bar{15}}s} & [_{88}^{\rm pv}]^{14}_ {\bm{\bar{15}}s} & [_{88}^{\rm pv}]^{15}_ {\bm{\bar{15}}s} & [_{88}^{\rm pv}]^{16}_ {\bm{\bar{15}}s} & [_{88}^{\rm pv}]^{17}_ {\bm{\bar{15}}s} & [_{88}^{\rm pv}]^{18}_ {\bm{\bar{15}}s} \\
\hline
\hline
 0  &  \sml{D^+\to \eta_8 \rho^+}  &  -\sqrt{\frac{3}{2}} \lambda   &  0  &  -\frac{\lambda }{\sqrt{6}}  &  -\frac{\lambda }{\sqrt{6}}  &  \frac{ \lambda}{2} \sqrt{\frac{3}{2}}
     &  -\frac{\lambda }{\sqrt{6}}  &  -\frac{\lambda }{\sqrt{6}}  &  -\sqrt{\frac{2}{3}} \lambda   &  -\frac{\lambda }{\sqrt{6}}  &  \frac{\lambda }{2 \sqrt{6}}
    &  -\frac{\lambda }{\sqrt{6}}  &  -\sqrt{\frac{2}{3}} \lambda   &  \sqrt{\frac{2}{3}} \lambda   &  0  &  -\frac{\lambda}{2} \sqrt{\frac{3}{2}}   &  -\frac{\lambda }{2
   \sqrt{6}}  &  -\frac{\lambda }{2 \sqrt{6}}  &  -\frac{\lambda }{2 \sqrt{6}}  \\
\hline
\end{tabular*}
}
}
\end{center}
\caption{$D^+ \to P_8V_8$ and $D^+_s \to P_8V_8$ $\mathcal{O}(\varepsilon)$ invariants generated by $\bm{\bar{15}}$.}
\label{tab:PV88S15DP}
\end{table} 

\clearpage


\end{document}